\documentclass{ar-1col}
\usepackage{cite}
\usepackage{soul,upgreek}
\usepackage{hyperref}        %%%% for arXiv
\usepackage{fullpage}        %%%% for arXiv ??

\jname{Annu. Rev. Nucl. Part. Sci.} \jvol{68} \jyear{2018}
\doi{10.1146/annurev-nucl-101916-123209}

\usepackage{xspace} % Include xspace

\newcommand{\ampt}{{\sc ampt}\xspace}

\newcommand{\bamps}{{\sc bamps}\xspace}
\newcommand{\vini}{{\sc vini}\xspace}
\newcommand{\mpc}{{\sc mpc}\xspace}
\newcommand{\zpc}{{\sc zpc}\xspace}
\newcommand{\superSONIC}{super{\sc SONIC}\xspace}

\newcommand{\mfp}{\ell_\mathrm{mfp}}
\newcommand{\pp}{\mbox{${p}+{p}$}\xspace}
\newcommand{\pau}{\mbox{$p+$Au}\xspace}
\newcommand{\pa}{\mbox{$p+A$}\xspace}
\newcommand{\ppb}{\mbox{$p+$Pb}\xspace}
\newcommand{\dau}{\mbox{$d+$Au}\xspace}

\newcommand{\nucnuc}{\mbox{$A+A$}\xspace}
\newcommand{\heau}{\mbox{$^3$He$+$Au}\xspace}

\newcommand{\pdhea}{\mbox{$p(d,^{3}$He)$+$$A$}\xspace}
\newcommand{\auau}{\mbox{Au$+$Au}\xspace}

\newcommand{\pbpb}{\mbox{Pb$+$Pb}\xspace}

\newcommand{\pt}{$p_{T}$\xspace}
\newcommand{\sqsn}{$\sqrt{s_{NN}}$\xspace}

\begin{document}

\markboth{Nagle $\bullet$ Zajc}{Small System Collectivity}

\title{Small System Collectivity in Relativistic Hadronic and Nuclear Collisions}

\author{James L.~Nagle$^1$ and William A.~Zajc$^2$
\affil{$^1$Department of Physics, University of Colorado, Boulder, Colorado 80309, USA;
email: jamie.nagle@colorado.edu} \affil{$^2$Department of Physics, Columbia University,
New York, NY 10027, USA; email: waz1@columbia.edu}}

\begin{abstract}
The bulk motion of nuclear matter at the ultrahigh temperatures
created in heavy ion collisions at the Relativistic Heavy Ion
Collider and the Large Hadron Collider is well described in terms
of nearly inviscid hydrodynamics, thereby establishing this system
of quarks and gluons as the most perfect fluid in nature. A
revolution in the field is under way, spearheaded by the discovery
of similar collective, fluid-like phenomena in much smaller
systems including \pp, \pa, \dau, and \heau collisions. We review
these exciting new observations and their profound implications for
hydrodynamic descriptions of small and/or out-of-equilibrium systems. 
\end{abstract}


\maketitle

\tableofcontents

\clearpage

\section{INTRODUCTION}

The modern era of heavy ion physics began in the year 2000, with
the first data taking at the Relativistic Heavy Ion Collider (RHIC),
followed by heavy ion running at the Large Hadron Collider (LHC)
in 2010. Shortly thereafter, a new arena for studying high-temperature nuclear matter came to the fore with a host of
observations in small collision systems, including \pp and \pdhea
collisions. These smaller systems exhibited many of the features
of collective behavior found in collisions of heavy nuclei
attributed to the perfect liquid 
%%%% Comment
%\textbf{\textcolor{red}{[**AU: Please note that quotation marks used to indicate nonstandard usage or irony have been removed, per house style**]}}
%\textbf{\textcolor{blue}{[So noted, we also changed 'perfect fluid' to 'perfect liquid']}}
%%% Comment
nature of quark--gluon plasma
\footnote{For an introduction to quark-gluon plasma properties, see the overview article that appears in this volume~\cite{Busza:2018rrf}.}
(QGP). While these observations were contrary to expectations,
there is a long history of considering even small collision
systems in the framework of hydrodynamics. This article begins
with a brief review of that history, followed by an overview of
observations from collisions of large nuclei such as gold and lead (\auau and \pbpb),
supporting the standard hydrodynamic model of heavy ion reactions.
Next we highlight the most important observations in small
collision systems that provide evidence for similar underlying
physics. We discuss key additional considerations
and alternative explanations. Finally we review the current status
of the theoretical interpretation of these results.

Before proceeding, a word on nomenclature is in order. 
In this review, we use ``collective'' as a generic descriptor for correlated particle production. If  $P(\vec{p}_1)$ is the probability to produce a particle with momentum $ \vec{p}_1$ in a collision, there is collective behavior if 
$P(\vec{p}_1 ), \vec{p}_2) \neq P(\vec{p}_1) P(\vec{p}_1) $. The nature of this correlation may be strictly at the two-particle level (e.g., resonance decay), or may extend to a broad number of particles, as in the case of jet production and (potentially) quantum interference effects. Hydrodynamic motion of a composite medium satisfies this definition of collective behavior, both at the two-particle and many-particle level.
We emphasize that these terms apply to observations, while the real physics questions lie in understanding the causes; hadronization in the case of jet production, inter-particle interactions and/or fields in the case of hydrodynamics. 

\section{HISTORICAL PRELUDES}\label{Sec:History}

Collective models of nuclear matter have a long history that is
replete with controversy and in some cases rancor. Certainly
Bohr's~\cite{Bohr:1936zz} compound nucleus and the associated
liquid-drop model (building on work by Gamow, Heisenberg, and von
Weizs\"{a}cker)~\cite{stuewer1994origin} evoked little
controversy, particularly after its quantitative success in
explaining fission in $^{235}\mathrm{U}$~\cite{Bohr:1939ej}. The
underlying physical assumption of quasi-equilibration of energy
was plausible when applied to reactions involving slow neutrons.
Collective descriptions of higher-energy collisions appeared to
be less well grounded. Heisenberg's~\cite{Heisenberg:1949kqa} 1949
attempt to understand excitations of the
pion fluid was widely ignored. By contrast, a year later
Fermi's~\cite{Fermi:1950jd} statistical model (which he
acknowledged to be the extreme limit of Heisenberg's approach)
received considerable attention. Fermi argued that precisely
because the interactions between pions and nucleons were strong,
one could expect the available energy to be ``rapidly \dots
distributed among the various degrees of freedom according to
statistical laws.''
%%% Comment
%\textbf{\textcolor{red}{[**AU: Please provide a page number for this direct quotation, if available**]}} 
%\textbf{\textcolor{blue}{[p. 570. This quote is from the first page of the corresponding journal article.]}}
%%% Comment
Taking the reaction volume as his only free
parameter, Fermi developed predictions (in modern terminology) for
particle multiplicities in terms of $n$-body phase space, and
presented a simple argument showing that in the high-energy limit
the number of produced particles $N$ would vary with
center-of-mass collision energy $\sqrt{s}$ as $N \sim s^{1/4}$.

Fermi carefully qualified his statistical model's extreme
assumptions, noting that by working in the opposite regime from a
perturbative approach, one might be able to bracket the correct
theory. As noted by Anderson in Fermi's {\em Collected
Papers}~\cite{fermi1962collected}, ``In the later literature this
made it appear that this theory was always wrong; a point that
Fermi didn't enjoy at all.''
%%% Comment
%\textbf{\textcolor{red}{[**AU: Please provide a page number for the two direct quotations in this paragraph, if available**]}} 
%\textbf{\textcolor{blue}{[For Reference 6, it's page 789. For Reference 7, it's page 51 (first page of the article)]}}
%%% Comment
A special case of such criticism was
voiced by Landau~\cite{Landau:1953gs}, who, after noting Fermi's
``ingenious idea,'' writes that ``the quantitative
calculation given by him appears unconvincing to us and incorrect
at several points.'' In particular, Landau observes that the
number of particles in the strongly interacting initial state is
ill-defined [a point he attributes to
Pomeranchuk~\cite{Pomeranchuk:1951ey}], and that the distribution
of final-state particles may be calculated only at the endpoint of
a {hydrodynamic}
%%% Comment
%\textbf{\textcolor{red}{[**AU: Please note that italic font used for emphasis has been removed, per house style**]}} expansion.
%\textbf{\textcolor{blue}{[OK, but this weakens the sentence; it's important for the reader to know that the expansion is hydrodynamic.}]}
%%% Comment
Landau~\cite{Landau:1953gs} also stated that the
hydrodynamic motion would be that of ``an ideal (non-viscous and
non-heat-conducting) liquid.'' In subsequent research, Landau \&
Belenkij~\cite{Belenkij:1956cd} elaborated on this
assertion, noting that the condition for the applicability of
hydrodynamics $ R \gg \mfp$, where $R$ is the least dimension of
the system and $\mfp$ is the mean free path, for a {relativistic} system necessarily leads to a large Reynolds number
\begin{marginnote}
\entry{Reynolds number}{The dimensionless ratio of inertial forces to dissipative forces in a fluid.}
\entry{Shear viscosity}{The larger the shear viscosity the more easily momentum can be exchanged between distant  fluid cells and, consequently, the faster a gradient in  fluid velocity (or a sound wave) dissipates into heat.}
\end{marginnote}
characteristic of inviscid systems. Expressing the Reynolds number
in terms of the mass density $\rho$, the shear viscosity
$\eta$, the bulk velocity $V$ of the system, and the microscopic
velocity $v$ of its constituents, one has

\begin{equation}
\mathrm{Re} \equiv \frac{\rho R V}{\eta} \sim \frac{\rho R
V}{\rho\ \mfp\ v} \sim \frac{R c}{\mfp\ c} = \frac{R}{\mfp} \gg 1
\quad . \label{Eq:LandauRe}
\end{equation}
Therefore, intrinsically relativistic systems in the hydrodynamic
limit should have low (kinematic) viscosities.

Despite the pedigree of these early developments, it is fair to
say that the hydrodynamic approach never entered the mainstream of
hadronic physics in the second half of the twentieth century. Rather, a
variety of methods---phase-shift analyses, S-matrix, bootstrap,
and so forth---were investigated before QCD emerged as the underlying
field theory for the strong interaction in the 1970s. Not even
the excellent hydrodynamic description of inclusive hadron
rapidity distributions at Fermi National Accelerator Laboratory (FNAL) fixed target energies and the CERN
Intersecting Storage Ring (ISR)~\cite{Carruthers:1973ws} was able to gain traction against the
subsequent enthusiasm for QCD's clear predictions for perturbative
phenomena. One of the few exceptions to this general trend was
Bjorken's~\cite{Bjorken:1982qr} simple and hugely influential model of hydrodynamic
expansion in ultrarelativistic \nucnuc
collisions, which explicitly allowed for its
application to \pp collisions.

Motivated by Bjorken's predictions, there were experimental
searches for signatures of QGP formation in \pp and $\bar{p}+p$
collisions, including Tevatron experiments
E735~\cite{Alexopoulos:1993wt} and MiniMAX~\cite{Brooks:1999xy}.
No firm conclusions resulted from this program, in part due to the
non-comprehensive nature of these experiments; for example, MiniMAX
exclusively searched for disoriented chiral condensates (DCC). In
hindsight, the DCC searches serve as an important reminder that
when a region of disturbed vacuum eventually returns to the
normal vacuum via particle emission the final number of
hadrons may not be the relevant quantity to understand whether
collectivity or hydrodynamics is applicable at earlier times. In
the case of E735, baryon and strangeness modifications in high-multiplicity events were a possible QGP signature, but the experiment also found
explanations via autocorrelations between higher multiplicities
and larger numbers of gluon jets. Recent measurements of
strangeness enhancement of multi-strange baryons in \pp collisions
at the LHC have revived this important discussion~\cite{ALICE:2017jyt}.
The field pushed forward to study the collisions of the largest
nuclei at relativistic energies, first in the fixed target
programs at the Brookhaven Alternating Gradient Synchrotron (AGS) and CERN Super Proton Synchrotron (SPS), and ultimately with
the construction of RHIC and the LHC.

\section{STANDARD MODEL OF HEAVY ION COLLISIONS}\label{Sec:SMofHIC}

In 2001, early results from the RHIC program indicated that in
head-on \auau collisions at 200~GeV per nucleon
pair, the majority of the energy is deposited into a medium whose
expansion is well described hydrodynamically, that is, as a flowing
fluid~\cite{Adcox:2004mh,Adams:2005dq,Back:2004je,Arsene:2004fa}.
The hydrodynamic nature of the matter was eventually quantified in
terms of its shear viscosity, which turns out to be very close to
the conjectured smallest possible ratio of viscosity to entropy
density [${\eta}/{s} \geq {\hbar}/({4 \pi k_\mathrm{B}}) =
{1}/{4\pi}$ in natural units] of any
fluid~\cite{Kovtun:2004de,Son:2007vk}. This nuclear matter has a
starting temperature of order 350--400~MeV, or equivalently four
trillion Kelvin, and as such is composed of quarks and gluons no
longer bound into color-neutral hadrons such as protons and
neutrons. Subsequent measurements of \pbpb
collisions at the LHC~\cite{Muller:2012zq} at up to 5.02~TeV per
nucleon pair display a similar fluidity with the matter starting
at a higher initial temperature of order
400--600~MeV \cite{Adam:2015lda}. In both cases, \auau and \pbpb
collisions create a QGP that behaves as a nearly perfect fluid, that is, a fluid with ${\eta}/{s} \sim {1}/{4\pi}$.

Just as the Big Bang theory is the prevailing paradigm for the
time evolution of the early Universe, over the last 10 years
the nuclear physics community has developed a Little Bang theory
as the standard model for the time evolution of heavy ion
collisions (described in detail in
References~\citen{Romatschke:2009im,Heinz:2013th,Jeon:2015dfa,Jaiswal:2016hex,Song:2017wtw}).
The evolution can be broken into distinct epochs:
\begin{enumerate}

\item[1] The highly Lorentz-contracted nuclei collide with a very short traversal
time (\hbox{$\ll 1$~fm/\textit{c}}). Predominantly through interactions of gluons
in the nuclei, often described in terms of gluon fields, energy is
deposited into the newly created medium. The initial, very
inhomogeneous, distribution of deposited energy in the transverse
plane, perpendicular to the beam direction, is referred to as the
initial condition.

\item[2] The matter is initially out of equilibrium, and some time is
required for it to equilibrate. During this time the matter
expands at nearly the speed of light in the longitudinal direction
and begins to expand radially in the transverse plane. This is
often referred to as the pre-equilibrium stage.

\item[3] After the matter
is nearly equilibrated,\footnote{For most of a decade, since the
identification of the applicability of hydrodynamics to heavy ion
collisions, the predominant thinking has been that for
hydrodynamic calculations to be valid the system must be nearly
equilibrated at an early time of order $\tau \approx
0.5$--$2.0$~fm/\textit{c}. Thus the time before this point is referred to as
pre-equilibrium, and an entire sub-area of the field has been
devoted to the rapid equilibration puzzle, trying to answer
the question of how the system equilibrates so fast. However,
it was recently realized that the hot nuclear matter may
never come close to
equilibration~\cite{Florkowski:2017olj,Romatschke:2017vte,Romatschke:2016hle}
and that a different explanation justifies the applicability of
hydrodynamics, as we discuss in detail in
Section~\ref{HydroDiscuss}. In this picture, the separation of
stages 2 and 3 is really only hydrodynamization (the
point where hydrodynamics is applicable), and the naming of stage
2 as pre-equilibrium is misleading and should be simply
pre-hydrodynamization.} it is modeled via viscous
hydrodynamics using an equation of state from lattice QCD
calculations. Deviations from equilibrium are accounted for with
shear and bulk viscosity terms.
\begin{marginnote}
\entry{Lattice QCD}{Numerical solution of QCD using a space-time lattice.}
\entry{Polyakov loop}{A gauge-invariant quantity which is zero in the confining phase and non-zero in the deconfined phase of QCD.}
\entry{Bulk viscosity}{A measure of the internal resistance of a fluid to expansion or compression.}
\end{marginnote}

\item[4] The fluid cools to a temperature corresponding to the QGP
crossover transition \cite{Aoki:2006we} $T \approx 170$~MeV (as
determined by the inflection point of the Polyakov loop, roughly
equivalent to the confinement--deconfinement
transition)~\cite{Aoki:2009sc} and then breaks up into hadrons, as
most commonly modeled via Cooper--Frye
freeze-out~\cite{Cooper:1974mv}.

\item[5] The resulting hadrons scatter, both inelastically, until what
is called chemical freeze-out, and elastically, until kinetic
freeze-out, at which time they are assigned their final-state
momenta as measured experimentally.
\end{enumerate}

Sophisticated computer modeling of large numbers of individual
collisions follow each of these stages through to predictions for
final hadrons that are measured experimentally. As with
constraining properties of the early Universe, this field has
advanced to multiparameter Bayesian
analyses~\cite{Petersen:2010zt,Novak:2013bqa,Pratt:2015zsa,Bernhard:2016tnd,Moreland:2017kdx}
to extract key properties of the medium, such as $\eta/s$, and
to assess the correlated sensitivities of the extracted values to
different assumptions for the initial conditions.

\begin{figure}[!h]
\includegraphics[width=\textwidth]{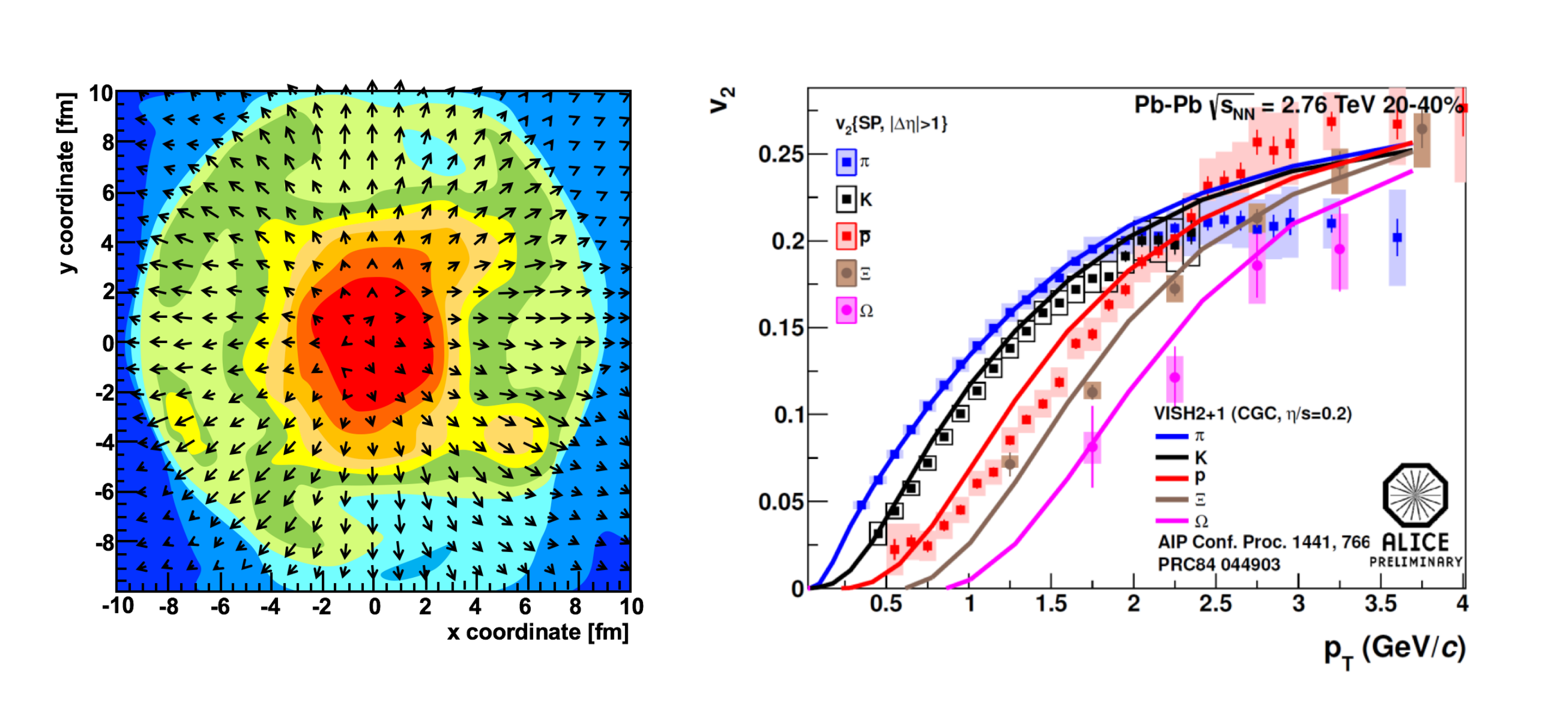}
\caption{(\textit{a}) Viscous hydrodynamic calculation results of a
semi-central \nucnuc collision in one time snapshot ($t =
5$~fm/\textit{c}). The color scale indicates the temperature of the fluid
cells in the transverse ($x$--$y$) plane, and the arrows represent
the fluid velocity vectors with the lengths proportional to the
speed. (\textit{b}) Elliptic flow coefficient $v_{2}$(\pt) for
identified hadrons in \pbpb collisions at the LHC, also compared
with hydrodynamic calculations~\cite{Shen:2011eg}.}
\label{fig:aa_flowprofile}
\end{figure}

The matter produced in the collision is subjected to enormous
longitudinal pressure, expanding at nearly the speed of light in
this direction, often assumed to be boost
invariant~\cite{Bjorken:1982qr}. There are also large pressure
gradients in the transverse direction driven not only by the
density differential to the vacuum outside the medium but also by
inhomogeneities in the matter. Figure~\ref{fig:aa_flowprofile}\textit{a}
shows the temperature and flow profile of an \nucnuc
collision from a hydrodynamic model. A number of key features are
worth describing in detail:

\begin{enumerate}
\item There is an overall pattern of strong radial outward expansion
with the largest bulk velocities near the periphery reaching 75\%
of the speed of light.

\item At the end of the hydrodynamic epoch, one calculates the hadronization
process in the rest frame of the fluid cell and then boosts
hadrons into the lab frame. Thus, heavier hadrons receive a larger
momentum shift (blue-shift) that is measurable as a distinct
feature in the transverse momentum (\pt) distribution of hadrons
as a function of their mass.

\item The spatial distribution of the matter and its temperature profile are
lumpy, despite the lumpiness of the initial condition already
having been washed out to some degree by viscous effects. These
inhomogeneities lead to substantial distortions in the azimuthal
distribution of particles~\cite{Ollitrault:1992bk}, which are
quantified in terms of a Fourier expansion~\cite{Voloshin:1994mz}
as

\begin{equation}
{{\mathrm{d}n} \over {\mathrm{d}\phi}} \propto 1 + \sum_{n} 2 v_{n}(p_T) \cos [n
(\phi - \Psi_{n})] \quad , \label{Eq:vn}
\end{equation}
where $p_T$ and $\phi$ are the transverse momentum and azimuthal
angle of each particle and $\Psi_{n}$ is the overall orientation
of the $n${th} moment. The first four moments, $v_{1}$, $v_{2}$,
$v_{3}$, and $v_{4}$, are often referred to as directed, elliptic,
triangular, and quadrangular flow coefficients, respectively.

\item Near mid-rapidity, for semi-central collisions, the dominant
\begin{marginnote}[-45pt]
\entry{Transverse momentum \pt}{The component of a particle's momentum $p$ that is transverse to the collision axis, $p_T \equiv p\ \cos \theta$, where $\theta$ is the particle's polar angle with respect to the collision axis.}
\entry{Central and off-central}{Ions colliding head-on are called central collisions, whereas if the ions only partially overlap the collision is semi-central or peripheral.}
%\entry{HBT correlations}{Correlations at low relative momentum between two identical particles that are sensitive to the size and shape of the emission region, named after Hanbury Brown (one person) and Twiss, who developed this method in astronomy.}
%\entry{Momentum rapidity y}{$\cosh(y)\equiv \gamma$, with $\gamma = 1/\sqrt{1-v_z^2}$, with $v_z$ the velocity along the beam direction in units of the speed of light.}
%\entry{Pseudorapidity $\eta$}{A mapping of the polar angle $\theta$ (with respect to the collision axis) of a particle, defined as $\eta = \frac{1}{2} \log \frac{1 + \cos \theta}{1 - \cos \theta}$. For massless particles pseudorapidity $\eta$ and momentum rapidity $y$ are identical; for highly relativistic particles rapidity and pseudorapidity are nearly equivalent.}
\end{marginnote}
Fourier coefficient is $v_2$, reflecting the efficient
hydrodynamic translation via pressure gradients of the initial
almond-shaped overlap region to momentum space. Again, because
of the larger fluid velocities built up along directions of
steeper pressure gradients, heavier hadrons will have their flow
patterns $v_n(p_{T})$ shifted outward in \pt (Figure~\ref{fig:aa_flowprofile}\textit{b}).
\end{enumerate}
Hydrodynamic calculations describe the measured higher-order
coefficients $v_3$ to $v_5$~\cite{Gale:2012rq}. Such comparisons
constrain both the initial inhomogeneities that are the source of
the fluid anisotropies and the medium properties such as the
shear viscosity, which has a larger damping effect on the
higher-order coefficients. There is a nice analogy between these
$v_{n}$ measurements in heavy ion physics reflecting the initial
spatial anisotropies and the spherical harmonic moment
measurements from the cosmic microwave background reflecting the
earlier inhomogeneities in the early Universe, providing key
constraints on QGP properties from the former and early Universe
properties from the latter.

The standard model of \nucnuc collisions has now been well established
and tested with great precision. This model describes a multitude
of experimental measurements including the mass-dependent \pt
spectra, the $v_{n}$ flow coefficients, the distribution of
event-by-event fluctuations in those flow coefficients,
multiparticle correlations referred to as
cumulants~\cite{Borghini:2000sa}, and Hanbury Brown Twiss (HBT)
correlations~\cite{Lisa:2005dd}. Other correlations between
\begin{marginnote}
\entry{HBT correlations}{Correlations at low relative momentum between two identical particles that are sensitive to the size and shape of the emission region, named after Hanbury Brown (one person) and Twiss, who developed this method in astronomy.}
\entry{Momentum rapidity y}{$\cosh(y)\equiv \gamma$, with $\gamma = 1/\sqrt{1-v_z^2}$, with $v_z$ the velocity along the beam direction in units of the speed of light.}
\entry{Pseudorapidity $\eta$}{A mapping of the polar angle $\theta$ (with respect to the collision axis) of a particle, defined as $\eta = \frac{1}{2} \log \frac{1 + \cos \theta}{1 - \cos \theta}$. For massless particles pseudorapidity $\eta$ and momentum rapidity $y$ are identical; for highly relativistic particles rapidity and pseudorapidity are nearly equivalent.}
\end{marginnote}
different flow coefficients that only arise from the nonlinear
mode mixing terms in hydrodynamics~\cite{Teaney:2012ke} are
qualitatively described, leading Heinz \&
Snellings~\cite{Heinz:2013th} to refer to this as an {{experimentum crucis}} in support of the hydrodynamic paradigm.
There are some outstanding puzzles that may turn out to be
reconciled within the standard framework [as was the case for the
so-called HBT puzzle~\cite{Pratt:2008qv}], or be the first hints of
additional physics. Specific examples include the flow moment
ordering in ultra-central collisions~\cite{Shen:2015qta} and
thermal photon emission and anisotropy~\cite{Adare:2015lcd}.

\section{SMALL SYSTEM EXPERIMENTAL DATA}

In the early years of the heavy ion collider era, small
colliding systems such as \pdhea were regarded as control
measurements. Measurements in \dau and \ppb collisions at RHIC and the LHC
have been very useful, for example, in constraining nuclear
modified parton distribution functions (nPDFs) that determine the
initial gluon distributions that determine the first epoch of
heavy ion collisions~\cite{Salgado:2016jws,Eskola:2009uj}.
However, in 2010, the CMS Collaboration examined ultrahigh-multiplicity \pp collisions at the LHC and found that particles
had a weak, though clear, preference to be emitted along a common
transverse $\phi$ angle across all
rapidities~\cite{Khachatryan:2010gv}. This finding sparked a scientific
debate over whether this could be related to similar correlations
observed in \nucnuc collisions, or was due to new physics coming from
momentum correlations present in the earliest moments of the collision.
Then in 2012, \ppb data
taking at the LHC, quickly followed by a reexamination of \dau
data at RHIC, revealed that most of the signatures for
hydrodynamic flow in \nucnuc collisions also existed in these smaller
systems. The revolution started by these small system
measurements, and the attempt to reconcile them in the context of
the heavy ion standard model, is the focus of this review. We
concentrate on those observables most directly related to
collectivity while noting that there is a wealth of data not
included on electroweak probes, strangeness enhancement, and so forth, and
additional physics areas of interest regarding nPDFs, gluon saturation phenomena,
multiparton interactions, and color reconnection, among others.

\subsection{Two-Particle Correlations and Initial Observations}

\begin{figure}[!h]
\includegraphics[width=\textwidth]{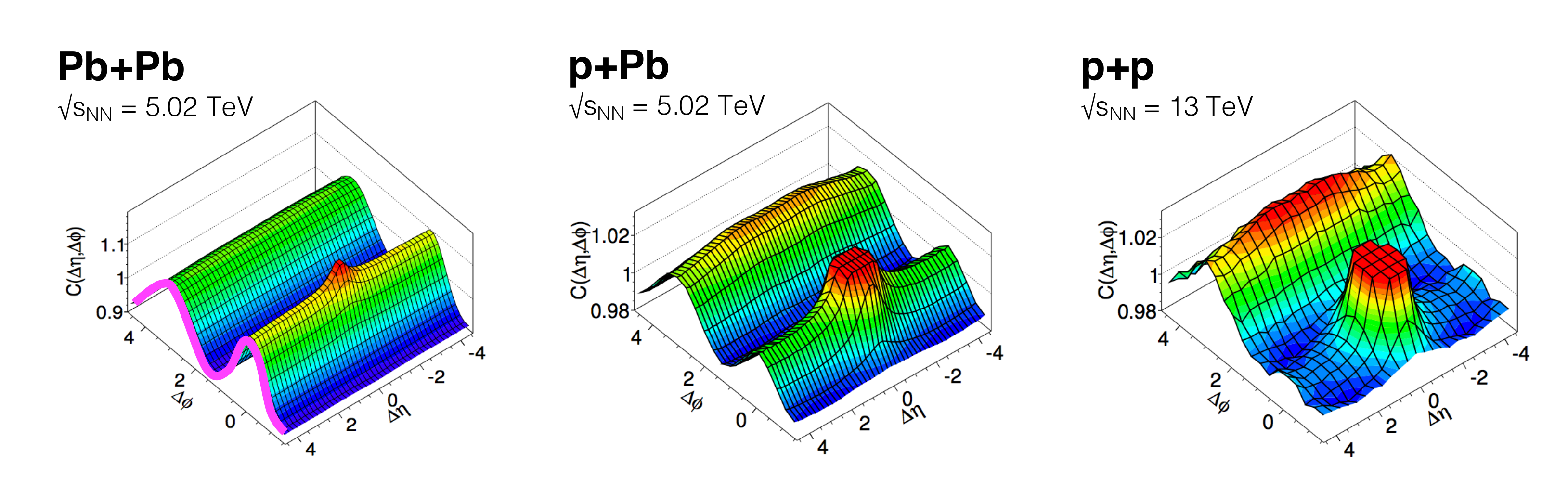}
\caption{Two-particle correlation results in (\textit{a})\ \pbpb , (\textit{b}) \ppb , and (\textit{c}) \pp  collisions at the
LHC~\cite{Aad:2015gqa}. In
\pbpb collisions there is a large cos$(2\Delta\phi)$ correlation
with peaks at $\Delta\phi = 0,\pi$ that extend long-range in
pseudorapidity $\Delta\eta$ (\textit{magenta curve}). A similar feature is observed in \ppb
and \pp collisions, thought it does not dominate the overall correlations
to the same degree.} \label{fig:threesystem_correlations}
\end{figure}

Crucial information regarding collectivity is garnered through the
measurement of two or more particle correlations, often
parameterized via the particles' relative azimuthal angle $\Delta
\phi$ in the transverse plane, and their relative longitudinal
pseudorapidity $\Delta \eta$. Since the reaction plane angles
$\Psi_n$ in Equation~\ref{Eq:vn} are assumed to reflect
geometric features of the initial matter distribution common to
all produced particles, standard Fourier properties lead to
two-particle correlations proportional to $v_n^2 \cos (n \Delta
\phi)$ that extend long-range in pseudorapidity as the matter
expands longitudinally. Figure~\ref{fig:threesystem_correlations}
shows two-particle correlations as a function of relative angles
$\Delta \phi$ and $\Delta \eta$ as measured in \pp, \ppb, and \pbpb
collisions at the LHC. In the \pbpb case, the long-range
correlations dominate and were originally referred to as the ridge
around $\Delta \phi = 0$ and another ridge around $\Delta \phi
= \pi$. In central collisions this second feature split into two
ridges near $\Delta \phi \approx 2\pi/3$ and $4\pi/3$ and for a
time were mistakenly interpreted as a Mach cone response from
high-energy quarks traversing the matter. These features are now
understood in a fully unified picture~\cite{Alver:2010gr} as
arising from elliptic, triangular, and higher flow moments.

However, there are a number of sources for such correlations
having nothing to do with a flowing medium. In a hydrodynamic
description, all of these other correlation sources are referred
to as non-flow. Simple examples include the decay of hadronic
resonances, such as $\Delta^{++} \rightarrow p+\pi^{+}$, giving rise
to a two-particle correlation. Large momentum-transfer scattering
of partons from the incoming hadrons or nuclei can result in jets,
that is, two collimated sprays of hadrons that are nearly
back to back in azimuth ($\Delta \phi \approx \pi$) and with a
correlation in pseudorapidity depending on the momentum fractions
$x_{1}$ and $x_{2}$ of the incoming partons. Even low momentum-transfer
scattering of initial partons can result in long-range
correlations in pseudorapidity as a consequence of total momentum conservation.
These contributions are evident in correlation measurements in all
collision systems from $e^{+}e^{-}$, \pp, and \nucnuc to varying
degrees and must be accounted for in order to isolate the
contribution from flow physics.

In the \pbpb case, in addition to the dominant flow contributions,
there is a localized peak near $\Delta \phi \approx \Delta \eta
\approx 0$ due to correlations among a small number of particles
from single jet fragmentation, resonance decay, and so forth. Because
hadrons from a single fragmenting jet are in a cone, they are
easily distinguished from the long-range flow contribution
around $\Delta \phi = 0$. However, the dijet partner, while
approximately back to back in azimuth, can swing in pseudorapidity,
resulting in a long-range correlation around $\Delta \phi =
\pi$, which is more challenging to disentangle from flow. In the
\nucnuc case, these dijet correlations are subdominant for \pt $<
5$~GeV/\textit{c}. Figure~\ref{fig:threesystem_correlations}
shows the same two-particle correlations in \ppb and \pp collisions
at the LHC. One observes the near-side ridge in both cases,
\begin{marginnote}
\entry{Near and away side}{Near-side refers to the region in azimuthal angle $\phi$ near a high transverse momentum particle or jet ``trigger''. Away-side is the region at relative azimuthal angle (with respect to the near-side trigger) $\Delta \phi \approx \pi$.}
\end{marginnote}
although with weaker strength, and a larger away-side ridge,
from the combination of flow correlations and non-flow
contributions. These features represent the first evidence of
flow-like collective behavior in a small system: High-multiplicity
\pp collisions at the LHC exhibit a long-range near-side ridge
in azimuthal correlations, very similar to that observed in
\nucnuc collisions. Because of the unexpected nature of the
ridge as a flow signature in small systems (though not
unexpected by all; e.g.,~\cite{Bozek:2011if}) and
the inability to determine whether there was a contribution on the
away side underneath the dijet signal, there was speculation of
possible new physics at play having nothing to do with the initial
geometry followed by collective expansion. We discuss these
alternative scenarios in Section~\ref{MomDomain}.

In 2012, \ppb collisions at $\sqrt{s_{NN}} = 5.02$~TeV were first
run at the LHC, and immediately all of the collaborations published similar
flow observations [see, e.g., results from
ALICE~\cite{Abelev:2012ola}, ATLAS~\cite{Aad:2012gla}, and
CMS~\cite{CMS:2012qk}]. Here the experimental signatures were much
stronger than in \pp collisions, and the race was on to repeat as
many of the \nucnuc measurements related to collectivity as possible to determine whether
the signals persisted in \ppb. Experimenters at RHIC immediately
reexamined \dau collision data at $\sqrt{s_{NN}} = 200$~GeV from
2008 and found similar patterns, though with a smaller flow signal
relative to the non-flow backgrounds~\cite{Adare:2013piz}. To
date, nearly all observations in \nucnuc collisions that provided
strong evidence for the heavy ion standard model ``quark--gluon plasma as
near-perfect fluid'' have now been measured in \ppb and \dau\ collisions
(see Reference~\cite{Loizides:2016tew} for an excellent review). The
\begin{marginnote}
\entry{Jet quenching}{The suppression of high transverse momentum particle and/or jet production relative to yields expected from the number of hard scatters in a collision.}
\end{marginnote}
notable exception to this statement is jet quenching, which is
discussed in Section~\ref{JetQuenching}.

\subsection{Instructive Measurements}

In this section we discuss four particularly instructive measurements in
small systems, each of which tests a key aspect of extending the
heavy ion standard model to such systems. These measurements
involve (\textit{a}) multiparticle cumulants demonstrating that
correlations exist among the majority of emitted particles as
opposed to a small subset, (\textit{b}) manipulation of the colliding small
nuclei to see whether the correlations scale as expected with initial
geometry, 
(\textit{c}) 
%%% Comment
%\textbf{\textcolor{red}{[**AU: OK\ to insert ``manipulation of'' here for parallelism?**]}}
%\textbf{\textcolor{blue}{[Yes, thank you.]}}
%%% Comment
particle-identified flow patterns to see whether they
reflect a common velocity field of a fluid at hadronization, and
(\textit{d}) higher moments of the flow patterns, including triangular and
quadrangular flow.

\subsubsection{Multiparticle cumulants}

In a collision creating $N$ particles, one can ask whether a given
two-particle correlation is indicative of correlations involving
only a small subset of particles $M \ll N$ (as in the dijet case),
or from $M \approx N$, that is, a feature of the bulk. Most
non-hydrodynamic explanations for the observations in small
systems invoking finite-size momentum domains predicted the former
case, whereas an overall flowing medium implies the latter case.
Multiparticle cumulants utilize sets of 2, 4, 6, ...,
$n$ particles that sequentially subtract away correlations
among only $n-2$ particles, with an extension to all $N$
particles using the Lee--Yang zeros
method~\cite{Bhalerao:2003xf,Borghini:2004ke}. These measurements
have been particularly powerful because in the small-variance
Gaussian limit the two-particle and four-particle results can be
written as $v_{2}\{2\} = \sqrt{\overline{v_{2}}^{2} + \sigma^{2}}$
and $v_{2}\{4\} \approx \sqrt{\overline{v_{2}}^{2} - \sigma^{2}}$.
They therefore allow extraction of both the event-averaged
$\overline{v_{2}}$ and the event-by-event variance
$\sigma^{2}$~\cite{Ollitrault:2009ie}. This has established in
\nucnuc collisions at RHIC and the LHC a direct quantitative
connection between the event-by-event variation in the initial
geometry and the flow fluctuations\cite{Giacalone:2017uqx}.

\begin{figure}[!h]
\includegraphics[width=\textwidth]{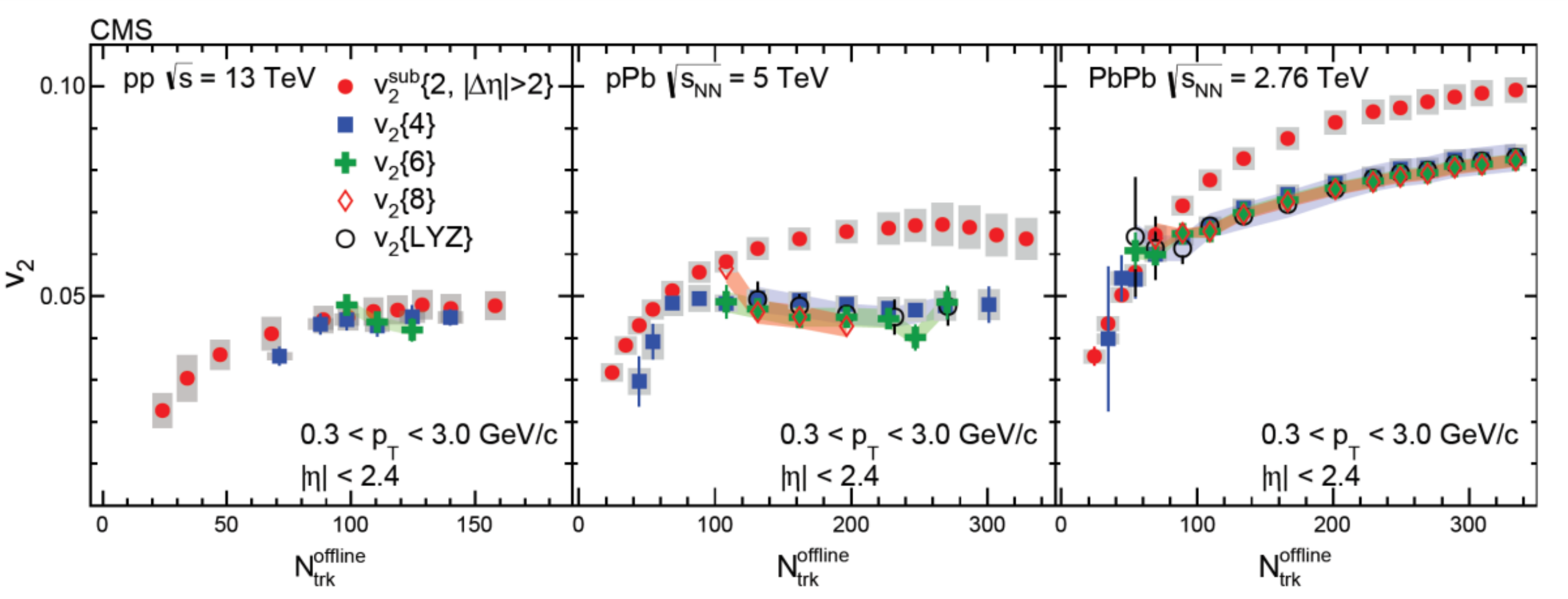}
\caption{The $v_{2}$ multiparticle cumulants as a
function of charged-particle multiplicity for (\textit{a}) \pp, (\textit{b}) \ppb, and (\textit{c}) \pbpb
collisions at the LHC~\cite{Khachatryan:2016txc}.}
\label{fig:cumulants}
\end{figure}

Figure~\ref{fig:cumulants} shows $v_{2}$ multiparticle cumulants
as measured in \pp, \ppb, and \pbpb collisions at the
LHC~\cite{Aad:2013fja,Chatrchyan:2013nka,Abelev:2014mda,Khachatryan:2015waa,Khachatryan:2016txc}.
The splitting of $v_{2}\{2\} > v_{2}\{4\}$, as related to flow
fluctuations, is also observed in \ppb collisions, yet disappears in
the \pp case. In 2016, RHIC had a special run of \dau collisions
over a range of energies (200, 62.4, 39, and 19.6 GeV) to address how
low in energy these features persist. The results from the
 \dau collisions at 200 GeV on the two-, four-, and six-particle cumulants also indicate that the
correlations are at the multiparticle
level~\cite{Aidala:2017ajz}.

We note that nonzero multiparticle cumulants are not unique to a
hydrodynamic description (e.g., \citen{Loizides:2016tew,Dusling:2017aot}). Imagine a flock of birds
in flight that have $N$-body correlations, where

\begin{extract}
order can be the effect of a top-down centralized control
mechanism (for example, due to the presence of one or more
leaders), or it can be a bottom-up self-organized feature emerging
from local behavioral rules. The prominent difference between the
centralized and the self-organized paradigm is not order, but
response~\cite{Cavagna29062010}.
%%% Comment
%\textbf{\textcolor{red}{[**AU: Please provide a page number for this direct quotation, if available**]}}
%\textbf{\textcolor{blue}
%{[p. 11865, the first page of the article.]}}
%%% Comment
\end{extract}
Thus, the key connection is the relation of cumulants to the
response to initial geometry, rather than the mere real-valued
\footnote{The $v_n$'s extracted using cumulants can assume complex values when large fluctuations and/or non-flow effects dominate the flow signal.}
$v_{2}, v_{4}, v_{6}$, and so forth.

In summary, the multiparticle measurements in \nucnuc, \ppb, and
\dau collisions at RHIC and the LHC yield strong evidence for
$N$-body correlations, providing a connection to the fluctuating
initial conditions. However, in the lower-multiplicity cases of
\pp at the LHC and \pau and lower-energy \dau at RHIC, the cumulants
do not follow the expected small variance expectation, which may
not be surprising as the fluctuations and non-flow effects are
larger. More research will be needed (e.g., \cite{Jia:2017hbm}) to
resolve these questions.

\subsubsection{Manipulating the geometry}

\begin{figure}[!ht]
\includegraphics[width=\textwidth]{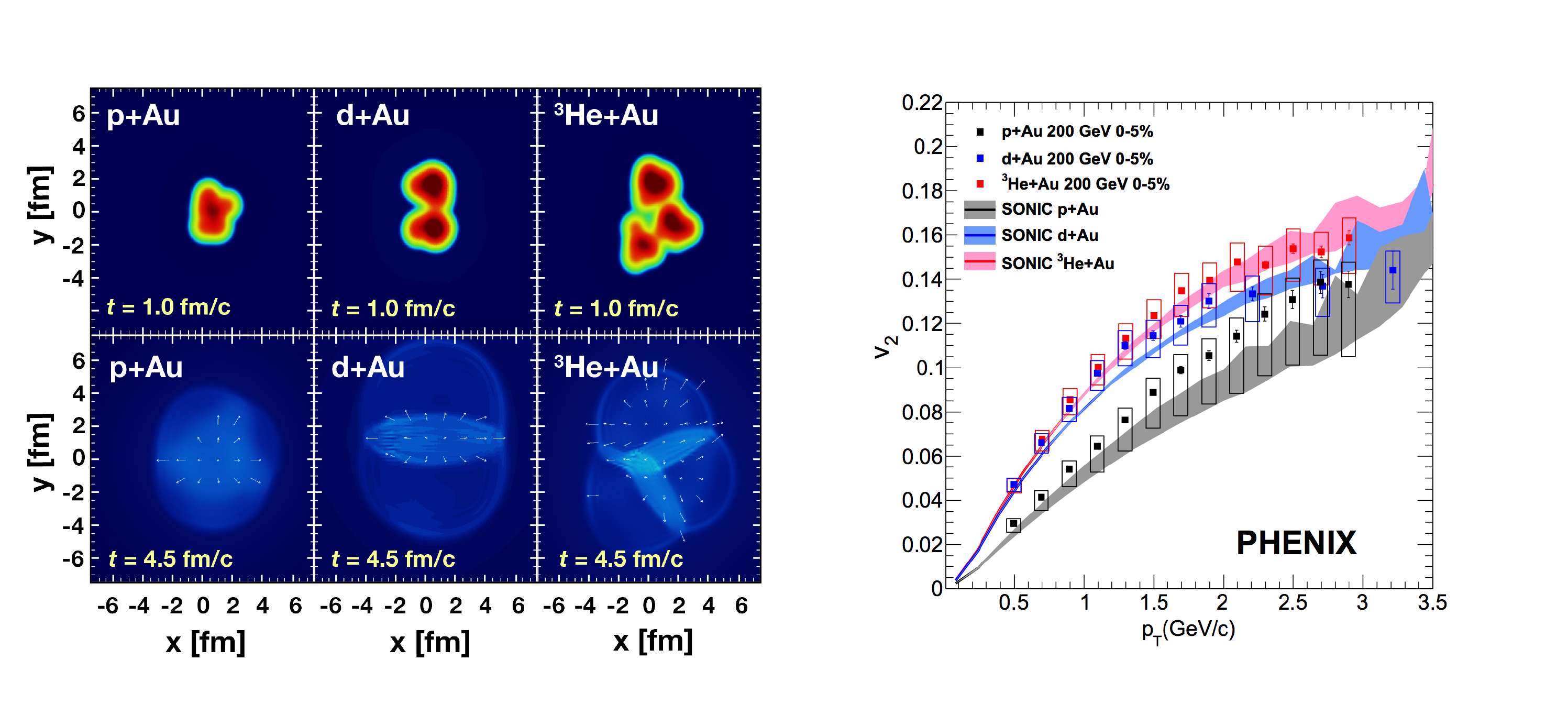}
\caption{(\textit{a}) Calculations of (\textit{top}) the initial energy density
in \pau, \dau, and \heau collisions at RHIC and (\textit{bottom}) the resulting
hydrodynamic evolution utilizing Monte Carlo Glauber initial conditions~\cite{Nagle:2013lja}. (\textit{b})
Comparison between hydrodynamic calculations~\cite{Nagle:2013lja} and
data from \pau, \dau, and \heau collisions at
$\sqrt{s_{NN}}=200$~GeV~\cite{Aidala:2016vgl}.}
\label{fig:rhicgeom}
\end{figure}

The initial small system flow measurements at RHIC were made in
\dau ~\cite{Adare:2013piz} rather than \pau\ collisions, due to
accelerator constraints. However, it was noted that in a \dau
central collision, the projectile neutron and proton from the
deuteron deposit energy in two hot spots, thus yielding a very
different initial condition than the single hot spot in a \pau
collision~\cite{Adare:2013nff}. This key
observation~\cite{Nagle:2013lja} led to a systematic program of
injecting different initial-state asymmetries through $p$,
$d$, and $^{3}$He projectiles incident on Au nuclei at
RHIC~\cite{Adare:2014keg,Adare:2015ctn,Aidala:2016vgl}. Figure~\ref{fig:rhicgeom}\textit{a} shows that the various projectiles
result in initial conditions that are dominantly circular,
elliptical, and triangular for \textit{p}, \textit{d}, and $^{3}$He projectiles,
respectively. Figure~\ref{fig:rhicgeom}\textit{b} shows theoretical predictions from the hydrodynamic standard
model~\cite{Nagle:2013lja} that are in excellent agreement with
the subsequent experimental measurements of $v_2$. In addition,
the $^{3}$He projectile was chosen to enhance triangular initial
geometries, and the triangular flow $v_{3}$ has also been measured
and is in agreement with theoretical
predictions~\cite{Adare:2015ctn}.

The agreement with data requires a full modeling of both the
initial conditions and the subsequent evolution. In the case of
\dau and \heau, the initial geometry is dominated by the location
of the two or three nucleons at the point of impact. In contrast,
for \pau, \ppb, and in particular \pp collisions, the initial
geometry depends critically on the modeling of sub-nucleonic
degrees of freedom (discussed in Section~\ref{ICsection}).
The simultaneous description of the three engineered geometries at
RHIC yields compelling evidence that the dominant correlation
source can be related to initial geometry coupled with subsequent
interactions or fluid dynamics. As of 
early 2018,
%\hl{this writing} 
%%% Comment
%\textbf{\textcolor{red}{[**AU: OK\ to replace with ``early 2018''?**]}}
%\textbf{\textcolor{blue}{[Yes; we've made the edit.]}}
%%% Comment
no alternative
explanation has been successfully put forward to describe these
observations.

\subsubsection{Mass-ordering fingerprint}

\begin{figure}[!h]
\includegraphics[width=\textwidth]{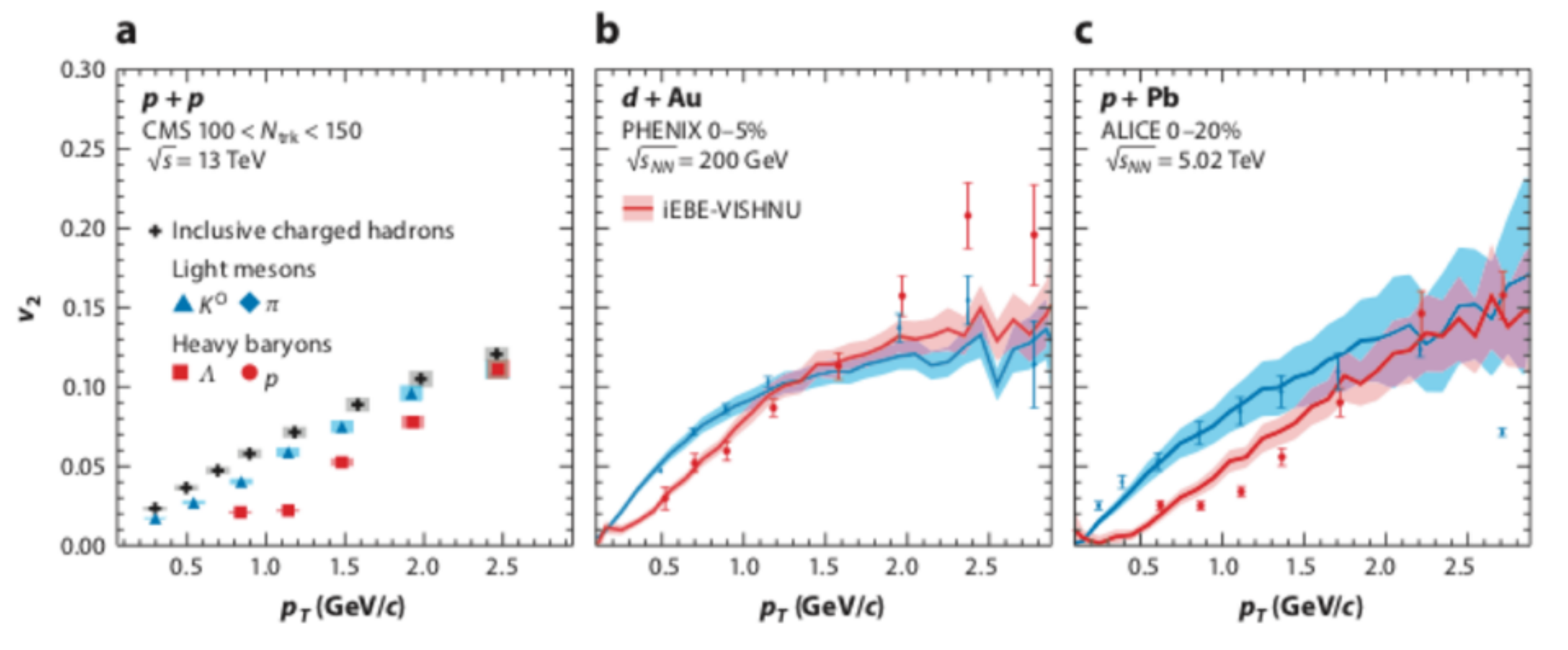}
\caption{Elliptic flow coefficient $v_2$ as a function of \pt for
different hadron species as measured in different small systems: (\textit{a})
\pp at the LHC, (\textit{b}) \dau at RHIC, and (\textit{c}) \ppb at the LHC. Theory calculations utilizing the hydrodynamic standard
model are from Reference~\cite{Shen:2016zpp}.}
\label{fig:massordering2}
\end{figure}

As noted in the \nucnuc case, there is a distinct ordering of
$v_{2}$ as a function of \pt for different hadron species. This
ordering is often referred to as the fingerprint of a flowing
fluid because it is the velocity of each fluid element as it
hadronizes that results in different momentum boosts for the
hadrons of different mass. Figure~\ref{fig:massordering2} shows
the mass dependence of $v_{2}$ in \pp~\cite{Khachatryan:2016txc},
\dau~\cite{Adare:2014keg}, and \ppb~\cite{ABELEV:2013wsa}
collisions, along with viscous hydrodynamic model comparisons in
the last two cases. The agreement between data and theory in the \dau
and \ppb cases at RHIC and LHC energies is another check on the
heavy ion standard model.

\subsubsection{Initial-state fluctuations and higher moments}

\begin{figure}[!h]
\includegraphics[width=\textwidth]{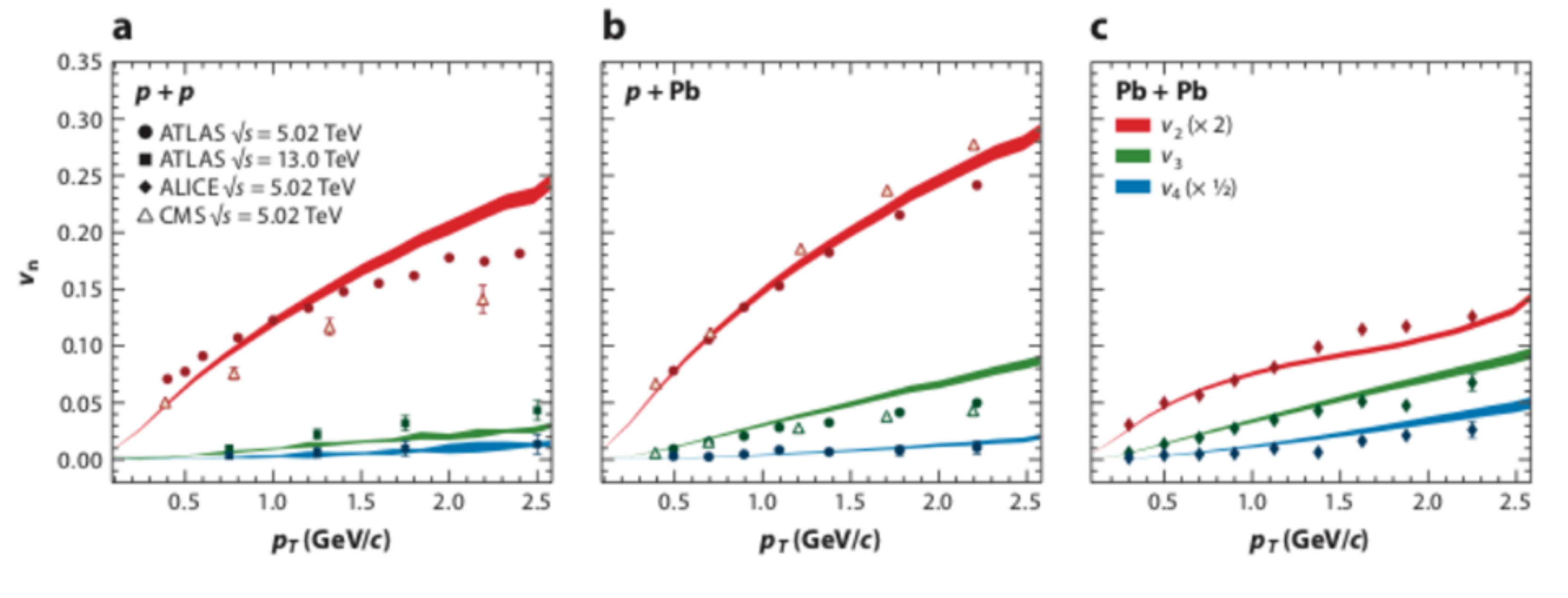}
\caption{Experimental data for momentum anisotropies $v_{2},
v_{3}$, and $v_{4}$ as a function of \pt in (\textit{a}) \pp, (\textit{b}) \ppb, and (\textit{c}) \pbpb collisions
at the LHC. Also shown are hydrodynamic standard model
(\superSONIC) calculations that incorporate constituent quark
Monte Carlo Glauber initial conditions, pre-equilibrium dynamics,
viscous hydrodynamics with $\eta/s = 1/4\pi$, and hadronic
scattering~\cite{Weller:2017tsr}.} \label{fig:sonicresults}
\end{figure}

A big step forward in solidifying the standard model for \nucnuc
collisions was the incorporation of nucleon-level fluctuations for
understanding the initial conditions and the resulting higher-order flow coefficients. For the \pp and \ppb cases, sub-nucleon-level
fluctuations are crucial, as discussed below in Section~\ref{ICsection}.
Figure~\ref{fig:sonicresults} shows the measured $v_{2}$, $v_{3}$,
and $v_{4}$ coefficients as a function of \pt in \pp, \ppb, and
\pbpb central collisions at LHC energies~\cite{Weller:2017tsr}.
Also shown are calculations from the SONIC implementation of the
heavy ion standard model starting with initial conditions based on
sub-nucleonic structure and $\eta/s = 1/4\pi$. Within the unified
framework of the heavy ion standard model, one achieves agreement
for all three systems and for all orders of $v_{n}$.

\subsection{Limits of Small System Flow Behavior}

All of these results engender the question: How low in deposited
energy, or in final particle multiplicity, does the experimental
data exhibit correlations that match viscous hydrodynamic
calculations? There are two different experimental ways to attack
this question: (\textit{a}) examining lower-multiplicity \pp interactions
and (\textit{b}) examining small system collisions at lower energy. After
the initial discovery of the \pp ridge at the LHC in
collisions at 0.9--7 TeV, follow-up measurements in \pp collisions
at higher energies of up to 13.1 TeV revealed an even stronger
signal. However, for \pp collisions of lower multiplicity, the
non-flow contributions increase and a reliable extraction of the
flow signal becomes model dependent. ATLAS~\cite{Aad:2015gqa}
and CMS~\cite{Khachatryan:2016txc} employ different extraction
methods and currently come to different conclusions regarding when
the flow signal disappears.

\begin{figure}[!h]
\includegraphics[width=\textwidth]{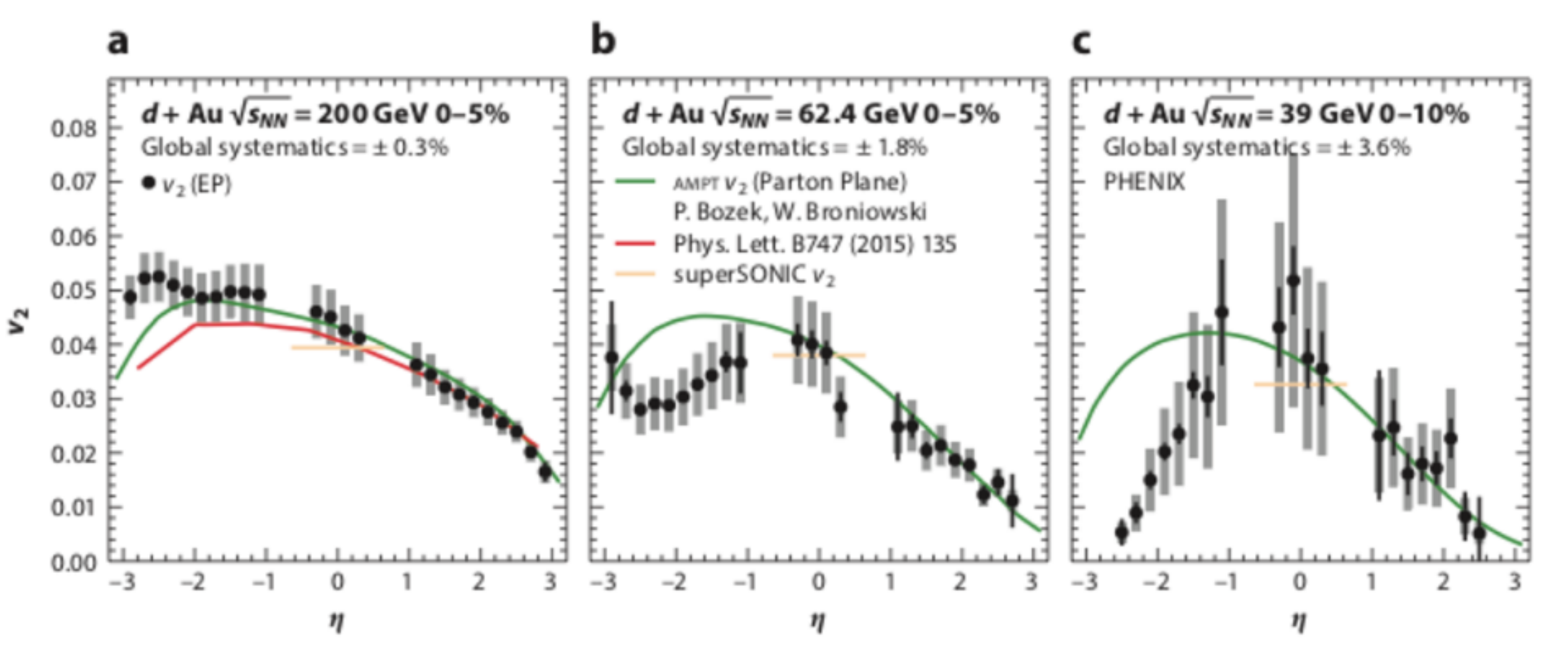}
\caption{Measured $v_{2}$ integrated over \pt as a function of
pseudorapidity from \dau collisions at (\textit{a}) 200, (\textit{b}) 62.4, and (\textit{c}) 39 GeV.
Also shown are calculations within the standard hydrodynamic
framework (\textit{red and orange curves}), as well as calculations from
the parton and hadron transport framework \ampt(A\ Multi-Phase Transport Model).} \label{fig:daubes}
\end{figure}

The other way to pursue this question is with the \dau beam
energy scan at RHIC. In this case, one has better control over the
initial geometry while changing the energy deposition and the
total particle multiplicity, albeit with larger theoretical
\begin{marginnote}
\entry{Transport coefficients}{parameterize the relaxation rate of a system 
in response to a gradient taking it out of equilibrium.
Examples include the thermal conductivity, electrical conductivity, and shear and bulk viscosities.}
\end{marginnote}
uncertainties due to the unknown variation of the transport
coefficients and the equation of state with the increasing baryon
chemical potential. Calculations within the hydrodynamic framework
predicted a rather modest decrease in the flow
signal~\cite{Romatschke:2015gxa,Koop:2015trj}. The
PHENIX Collaboration~\cite{Aidala:2017pup} has reported results on $v_{2}$ as a function of collision energy
(200, 62.4, 39, and 19.6 GeV). Figure~\ref{fig:daubes} shows the
measured $v_{2}$ coefficients as a function of pseudorapidity for
high-multiplicity \dau collisions at the three higher energies.
The measured $v_{2}$ shows little energy dependence, in reasonable
agreement with hydrodynamic calculations. Also shown are parton
transport model calculations that are described in detail in
Section~\ref{PartonScatter}. In \dau central collision data at 200
GeV, as noted above, there is additional evidence from the two-,
four-, and six-particle cumulants that the anisotropy is a bulk
$N$-particle correlation dominated by the translation of initial
geometry into momentum space. The flow signal via cumulants
appears to persist down to the lowest energies measured, though
masked by a growing non-flow contribution to the correlations.
The question of how small or low in energy these collective
features persist remains outstanding, and its resolution may hinge
on whether one can perfectly factorize the flow and non-flow
contributions.

\section{ADDITIONAL CONSIDERATIONS}

Several additional considerations are important
to include in any discussion of small system heavy ion physics.
Here we discuss two key topics that must be reconciled when
applying the heavy ion standard model to these small systems: (\textit{a})
the apparent absence of jet quenching effects in small collision
systems and (\textit{b}) the influence of modeling the initial conditions at
the sub-nucleonic level. In addition, there are proposed
alternative interpretations of the small system data that include
\begin{marginnote}
\entry{quasiparticles}{quasiparticles are excitations in a system with lifetimes long compared to the mean time between collisions, thereby allowing for a particle-like description of their transport properties.}
\end{marginnote}
(\textit{a}) parton scattering models with well-defined quasiparticles and
(\textit{b}) initial-state momentum correlation models. We discuss these
considerations in detail in the next four subsections.

\subsection{Jet Quenching in Small Collision Systems? \label{JetQuenching}}

\begin{figure}[!h]
\includegraphics[width=\textwidth]{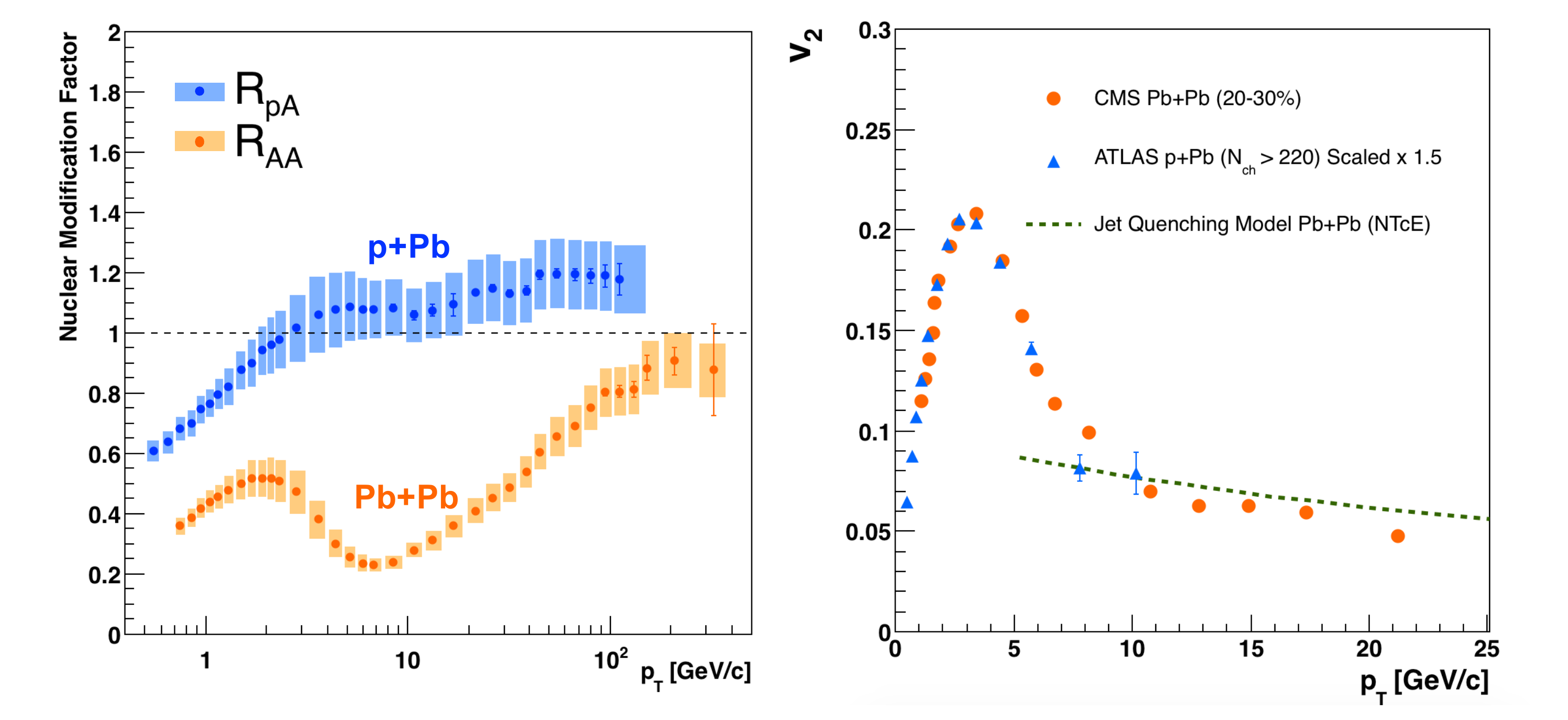}
\caption{(\textit{a}) The nuclear modification factor $R_{AA}$ for
unidentified hadrons as a function of
\pt~\cite{Khachatryan:2016odn}. This factor is the ratio of yields
in \nucnuc collisions relative to scaled up \pp yields. One
observes significant modification, namely suppression, in \pbpb
collisions and almost no modification in \ppb collisions. (\textit{b})
The $v_{2}$ coefficient for hadrons as a function of \pt in \pbpb
and \ppb collisions at the LHC. The \ppb results have been scaled
up by a factor of 1.5 for shape comparison. 
The 
%theory
green dotted
curve
%%% Comment
%\textbf{\textcolor{red}{[**AU: The green curve?**]}}
%\textbf{\textcolor{blue}{[Yes; we've edited the text.]}}
%%% Comment
~\cite{Zhang:2013oca} is from a jet quenching calculation
where the anisotropy results from the directional dependence of
the energy loss, rather than hydrodynamic flow.}
\label{fig:nojetquenching}
\end{figure}

In \nucnuc collisions, an important confirmation of the heavy ion standard
model comes from the energy loss of high-\pt partons traversing
the medium, referred to as jet
quenching~\cite{Gyulassy:2003mc,Norbeck:2014loa,Accardi:2009qv}.
Jet quenching models calculate the rate and kinematics for hard
scattering, that is, large momentum-transfer interactions, and then
propagate the resulting partons through the space-time evolution
of the matter calculated from hydrodynamic codes. Jet quenching
was discovered at RHIC in \auau collisions as a factor-of-five
suppression of high-\pt hadrons relative to their expected rate
from scaling up \pp yields~\cite{Adcox:2001jp}. 
%\hl{A critical measurement in 2003 was to see this ``quenching'' effect disappear}
A\ critical observation made in 2003 was that this quenching effect disappeared
%%% Comment
%\textbf{\textcolor{red}{[**AU: OK\ to reword as ``A\ critical observation in 2003 was that this quenching effect disappeared''?**]}}
%\textbf{\textcolor{blue}{[Yes, we've made this change.]}}
%%% Comment
in \dau collisions where no dense medium was
expected~\citen{Adler:2003ii,Adams:2003im,Arsene:2003yk,Back:2003ns}. No suppression was
observed in \dau collisions; thus, at the time, jet quenching
was confirmed as an exclusively final-state effect from the medium
in \nucnuc collisions. 
Similar measurements at the LHC of single
hadrons in \pbpb and \ppb collisions (Figure~\ref{fig:nojetquenching}\textit{a})~\cite{Khachatryan:2016odn}
demonstrate the quenching
observed in \nucnuc collisions is not observed in small systems.
Modern measurements including fully reconstructed jets and jet
structure provide further evidence for quenching-related
modifications in \nucnuc  but not in \pa collisions.

It now may seem surprising that no jet quenching effect is
apparent in \pa collisions if indeed a hot medium is formed. How
can there be a medium created that is described by hydrodynamics,
and that significantly modifies the distribution of final-state
hadrons, yet has no significant impact on the distribution of high-\pt particles? The jet quenching effect in \nucnuc collisions
becomes more prominent in more central, higher-multiplicity
reactions as the average in-medium path the partons traverse
correspondingly grows. In small systems, the medium created is
smaller, so the average path is expected to be significantly
shorter. One possibility is that after the hard scattering the
parton is in a highly virtual state, and its evolution may be only
modestly affected by scattering with other partons in the medium.
As such, a long medium traversal time, as in central \nucnuc
collisions, encompasses a parton where the medium scattering
significantly modifies the parton shower, but in a \pa collision
with short medium lifetimes ($\tau < 2$--$4$~fm/\textit{c}) the jet
quenching may be much smaller.

Quantitative theoretical calculations of the expected quenching
effects in small systems have been made (e.g.,
\citen{Park:2016jap,Tywoniuk:2014hta,Zhang:2013oca}), though
no clear consensus on the magnitude of the quenching has been
reached. The lack of quenching observed in
Figure~\ref{fig:nojetquenching}\textit{a} is observed in minimum
bias collisions---that is, averaged over all geometries. Many
calculations predict observable quenching effects in central or
high-multiplicity \ppb collisions where the paths traversed by the
partons may be longer. However, in these small systems, the
selection of event classes based on multiplicity has strong
autocorrelations between the nature of the nucleon--nucleon
collisions and the hard process itself, which complicate the
interpretation of experimental observables. When selecting on
multiplicity classes, jets and high-\pt hadrons are suppressed in
central events (as expected from jet quenching), but
are counterbalanced by an equal-magnitude enhancement in
peripheral events---thus resulting in no modification when
averaging over all \pa collisions~\cite{Perepelitsa:2017ltf}.
The suppression in central events is widely interpreted in terms
of this autocorrelation bias, that is, pushing more events with jets
into the most central category via multiplicity, rather than the
result of jet quenching~\cite{Adam:2014qja}. Recent results with
\begin{marginnote}
\entry{Spectator}{Nucleon that does not collide and hence keeps on moving along the beam direction.}
\end{marginnote}
event class selected on spectator neutrons, and thus with reduced
autocorrelation bias, indicate little or no quenching in more
central event categories~\cite{Adam:2016jfp}.

One mystery involves the measurement of $v_{2}$ for hadrons at
large \pt. In the \nucnuc case, the azimuthal anisotropy $v_{2}$
is interpreted in terms of flow for low-\pt particles. In contrast,
at high \pt, hadrons have a more modest anisotropy in \nucnuc
collisions (Figure~\ref{fig:nojetquenching}). This
anisotropy is thought to result from jet quenching, with partons
losing more energy when traversing a longer path through the
medium. What is striking is that the $v_{2}$ measured in
\ppb~\cite{Aad:2014lta} scaled by a factor of 1.5 (Figure~\ref{fig:nojetquenching}\textit{b}) appears to
follow the same pattern. If there is no jet quenching in \ppb
events, what else could be the source of the anisotropy at high
\pt?

Jet quenching is a fertile area of investigation and part of the
motivation for comparing a full suite of jet measurements at RHIC
from the new sPHENIX detector~\cite{Adare:2015kwa} to observations
at the LHC over a range of collision system sizes. New
measurements of charm and beauty hadrons in small systems are also
expected to be illuminating. In \nucnuc collisions, bulk medium
hydrodynamics and rare jet quenching probes provide complementary
information on the system created, and the presence of the former
and the apparent absence of the latter in small systems represent a crucial
area where more data and theoretical work are needed.

\subsection{Initial Conditions \label{ICsection}}

Extracting medium properties from hydrodynamic calculations requires a good quantitative constraint on the initial geometry.
The simplest such geometry is calculated via the billiard ball
interaction picture encapsulated in Monte Carlo Glauber
calculations~\cite{Miller:2007ri}. Within this framework,
individual nucleons are distributed within a nucleus following the
relevant Woods--Saxon functional form and the inclusion of a
hard-core repulsive potential. Nucleons in the projectile and
target then interact as dictated by the nucleon--nucleon inelastic
cross section. The resulting energy or entropy is distributed in
the transverse plane according to a two-dimensional Gaussian with
a width parameter typically chosen as $\sigma = 0.4$ fm.
Extensions to this picture incorporate nondeterministic
interaction probabilities, fluctuating nucleon sizes, and negative
binomial fluctuations in energy deposition (e.g.,
\citen{Welsh:2016siu,Broniowski:2007nz}).

In the case of \nucnuc collisions, the (extended) nucleon-level
Monte Carlo Glauber framework is, for the most part, sufficiently
constrained to provide confidence in the overall heavy ion standard model
space-time evolution and extraction of matter properties such as
$\eta/s$ with precision. This methodology was developed over many
years and includes refinements such as the inclusion of
deformation parameters, particularly for uranium, as well as
detailed studies of nucleon--nucleon correlations. In the case of
\dau and \heau, respectively, the Hulthen wave function for the
deuteron is well understood and for $^{3}$He the full three-body
wave function has been solved {{ab initio}}. These
calculations have shown that detailed substructure of the
nucleon influences results only in the most central \nucnuc
collisions, where fluctuations are the dominant source of azimuthal
anisotropies. However, in \pp and proton-induced nuclear
collisions, sub-nucleonic structure dominates and casts a shadow
over the predictive power of the standard heavy ion modeling.

Several studies of the influence of sub-nucleonic structure
modeling have been carried out,
%\cite{Noronha-Hostler:2015coa,Gelis:2016upa}, 
and we
describe here two such studies (see also \citen{Noronha-Hostler:2015coa,Gelis:2016upa}). The first utilizes the
IP-Glasma framework~\cite{Schenke:2012wb}, where the initial energy deposition is
computed in terms of overlapping gluon fields. In this calculation
the geometry of deposited energy follows the overlap regions
between interacting nucleons; therefore, in central \ppb and \pau
collisions it results in a very circular medium. 
\begin{marginnote}
\entry{Glasma}{A hypothesized precursor state to quark-gluon plasma where the color fields of the gluons form a classical coherent field with amorphous structure similar to glasses.}
\end{marginnote}
This circular
initial condition coupled with hydrodynamic evolution
underpredicts the $v_{2}$ in \pau and \ppb by up to a factor of four~\cite{Schenke:2014zha}. 
Thus, with very little initial eccentricity the medium
simply expands radially with no strongly preferred axis. A second
method utilizing the proton form factor~\cite{Habich:2015rtj} also
results in a very circular initial conditions and, in particular
for \pp collisions, predicts a vanishing $v_{2}$ for the highest-multiplicity collisions. 

%Models in which the initial proton has
%deformations that fluctuate event by event can reconcile the \ppb
%data, for example, but require additional parameters for each flow
%moment $v_{n}$~\cite{Schlichting:2014ipa} that potentially may be
%constrained from electron deep-inelastic scattering data.
%REMOVE THE ABOVE SENTENCE ENTIRELY.

Another approach is to include the simplest extension of the
sub-nucleonic picture by assuming the proton is decomposed into
three valence quarks, each with a cloud of gluons around it, and
that each valence quark cloud interacts when it comes within
some fixed distance of another such cloud. In much of the
literature this picture is referred to as the constituent quark
model. The hydrodynamic calculation shown in
Figure~\ref{fig:sonicresults} provides a reasonable description of
 $v_{2}$, $v_{3}$, and $v_{4}$ in \pp, \ppb, and \pbpb collisions
using this constituent-quark-based  Monte Carlo Glauber model for the initial geometry.
Further substructure, smaller than these clouds, is expected
to have a small influence, as a number of studies indicate that
finer-scale structures are very quickly washed out.

The earlier IP-Glasma calculations have also been extended under the ansatz that the proton has a substructure with three gluon hot spots and then constraining their distribution with one additional free parameter fixed to match HERA e+p data ~\cite{Mantysaari:2016ykx,Mantysaari:2016jaz}.   Applying this updated proton substructure as initial conditions in \ppb collisions, good agreement is found with many flow observables including higher moments~\cite{Mantysaari:2017cni}.

An intriguing new development involves inverting the problem: If one
posits viscous hydrodynamics as the correct model for the time
evolution, then one can try to determine the initial condition and
learn something about the structure of the proton on timescales
that are short compared with the nucleus crossing time \cite{Gelis:2016upa,Mantysaari:2017cni}.
%~\cite{Gelis:2016upa,Mantysaari:2017cni}. 
%QUESTION TO REMOVE THE REFRENCES ABOVE ENTIRELY...
At RHIC energies, this may be feasible because one can test the hydrodynamic evolution
hypothesis with \dau and \heau data that are not as sensitive to
the initial condition model. There have also been attempts to
simultaneously constrain medium properties and initial condition
substructure within a Bayesian
framework~\cite{Petersen:2010zt,Novak:2013bqa,Pratt:2015zsa,Bernhard:2016tnd,Moreland:2017kdx}.
This is an exciting prospect and should be fully pursued for small
system geometries at RHIC and the LHC.

\subsection{Parton Transport Models \label{PartonScatter}}

In the 1990s parton transport models were developed that treated
quarks and gluons as well-defined quasiparticles that scatter
with one another. Early implementations such as
\vini~\cite{Geiger:1991nj}, \zpc~\cite{Zhang:1997ej,Zhang:1999rs},
and \mpc~\cite{Molnar:2000jh} 
%%% Comment
%\textbf{\textcolor{red}{[**AU: Necessary to spell out model names here and in next paragraph (as for \ampt, below)?**]}}
%\textbf{\textcolor{blue}{[No, not really, as a) these are not discussed below and b) the models are almost always referred to by these acronyms.]}}
%%% Comment
predicted rather modest collective
effects (i.e., flow) due to the expected small QCD ($2 \rightarrow
2$) parton--parton scattering cross section. Had these calculations
proved accurate, the produced medium could accurately be termed a
weakly coupled QGP. Instead, the first experimental data from RHIC
\begin{marginnote}
\entry{\sqsn}{is the total collision energy per nucleon-nucleon pair in the center of mass frame.}
\end{marginnote}
with \auau collisions at \sqsn=130 GeV, indicating large elliptic
flow, immediately presented a major challenge for these frameworks.
For example, within \mpc, only by artificially increasing the
expected perturbative QCD parton--parton inelastic cross section from 3~mb to
45~mb could one describe the
data~\cite{Molnar:2001ux,Molnar:2003ff}. The conclusion at the
time was that the medium is strongly coupled; in other words, the
parton--parton interactions are highly non-perturbative, and there
are no well-defined quasiparticles. Thus, the system is amenable only
to calculations with strong fields or hydrodynamic
descriptors.

However, a new class of parton transport models has been developed
that provide a better qualitative description of experimental
data. Two such examples are the \bamps (Boltzmann Approach to Multi-Parton Scatterings)~\cite{Xu:2004mz} and \ampt (A Multi-Phase Transport)~\cite{Lin:2004en} models. The \bamps model considers only
gluon quasiparticles subject to $gg \rightarrow gg$ scattering
with a strong coupling $\alpha_{s} = 0.6$ as well as higher-order
scatterings of $gg \rightarrow ggg$ and $ggg \rightarrow gg$. In
the limit of many scatterings, \bamps produces hydrodynamic-like
flow patterns, and within this framework, a small effective
$\eta/s$ value near the lowest bound is
extracted~\cite{Xu:2007ns}.

The \ampt 
%%% Comment
%\textbf{\textcolor{red}{[**AU: OK\ to move definition to where model name first appears, above?**]}}
%\textbf{\textcolor{blue}{[Yes; we've edited the text.]}}
%%% Comment
model~\cite{Lin:2004en} has nearly
massless quark and antiquark quasiparticles that are produced
via a so-called string-melting mechanism. The produced quarks are
allowed to scatter, hadronize via coalescence, and then undergo
hadronic inelastic and elastic scattering. The implementation of
only quarks and antiquarks enables a consistent recombination
into hadrons via coalescence in the latter stage. This generator,
though with many tunable components and various seemingly
unphysical assumptions (e.g., no gluons), has successfully matched
a number of \nucnuc observables and provided insights into the
translation of initial geometry into final hadron momentum
anisotropies (most famously in Reference~\cite{Alver:2010gr}). It was
generally assumed that this was due to many scatterings that
effectively modeled fluid flow, that is, approaching the hydrodynamic
limit as the mean free path approaches zero in the transport
picture. However, recently it was shown that the number of
scatterings is quite modest, and for small systems at RHIC and the
LHC the majority of partons have no scatterings at all. This
realization has led to an understanding of the anisotropies as due
to a differential probability to scatter or not---a so-called
parton escape or tomographic image
scenario~\cite{He:2015hfa}. This puzzle is highlighted by the
agreement (often quantitative) between \ampt and small system flow
signals in \ppb~\cite{Bzdak:2014dia} and \pau, \dau, and
\heau~collisions \cite{Koop:2015wea}. Figure~\ref{fig:daubes} shows an example of this agreement with calculations of $v_{2}$ as a
function of pseudorapidity in \dau collisions at different
energies~\cite{Aidala:2017pup}.

An important set of outstanding questions includes the following: (\textit{a}) Are these
parton quasiparticle scattering scenarios a dual picture to
hydrodynamics even with very small scattering probability, and
(\textit{b}) if not, are there key distinguishing experimental observables
that can discriminate between the two? The latter question has
proven challenging to answer since many observables are sensitive to
the initial geometry and fluctuations, yet rather insensitive to
the mechanism of translation into momentum anisotropies. Thus,
\ampt describes  $v_{2}$, $v_{3}$, and $v_{4}$ and their
fluctuations at the same level as hydrodynamics when utilizing the
same initial conditions. Observables one would na\"ively expect to
be more sensitive, such as the mass-dependent $v_{2}$ splitting, are
in fact qualitatively reproduced in \ampt, yet result from
completely different physics---in this case from the hadronic
scattering stage~\cite{Li:2016ubw}. Another observable is the
correlation between flow moments, for example, $v_{2}$ and $v_{4}$,
that arise in hydrodynamics from nonlinear terms that result in mode
mixing. However, \ampt calculations achieve a similar level of
agreement with these mode-mixing observables~\cite{Yan:2015lwn}.

These models are seemingly self-contradictory. Both \ampt and
\bamps have a short initial formation time for the partons to
interact as well-defined quasiparticles and a mean free path
between scatterings shorter than the de~Broglie wavelength
$\lambda_\mathrm{dBg}$. In fact, in \ampt the initial parton formation
time is approximately 0.2 $\times~\lambda_\mathrm{dBg}$. Is it right to
then state that a weakly interacting system of partons is an
alternate picture to the strongly coupled hydrodynamics when
setting $\alpha_{s} = 0.6$ and assuming mean free paths less than
the de~Broglie wavelength ($\lambda_\mathrm{mfp} < \lambda_\mathrm{dBg}$)? It is
interesting to note that a precursor to the $\eta/s \ge 1/4\pi$
bound was indeed derived in kinetic theory under the assumption that a
particle mean free path 
%could not be shorter than the de~Broglie wavelength~\cite{Danielewicz:1984ww}. 
could not be smaller than the shortest distance resolvable via the uncertainty principle, corresponding to the reduced
de Broglie wavelength $\lambda_{dBg} / 2\pi$~\cite{Danielewicz:1984ww}. 
That said, the quantitative
description of a large collection of experimental data implies
that there is some key physics captured or mimicked in this
approach. The field requires a concentrated effort in developing
additional parton scattering models that are publicly available
(as is \ampt) that will simplify the physics assumptions to understand
how to reconcile or discriminate this quasiparticle picture from
the strongly coupled hydrodynamic one.

\subsection{Momentum Correlations Explanations \label{MomDomain}}

Both viscous hydrodynamics and parton transport calculations have
a common feature: The initial geometry of the deposited energy in
the transverse plane is translated into azimuthal momentum
anisotropies via final-state interactions, between either fluid
elements or quasiparticles. In contrast, when long-range
ridge correlations were first reported in high-multiplicity
\pp collisions at the LHC~\cite{Khachatryan:2010gv}, explanations
emerged in which the correlations were generated in the initial
scattering, that is, on the timescale of the nuclear crossing, and
required no later-stage interactions or coupling. A number of
these initial momentum correlation calculations are discussed in detail in
Reference~\cite{Dusling:2015gta}.

One proposal utilizes glasma
%%% Comment
%\textbf{\textcolor{red}{[**AU: Should glasma be capitalized here, as above?**]}}
%\textbf{\textcolor{blue}{[No; not clear why IP-Glasma authors chose to capitalize it but here it should be lower case.]}}
%%% Comment
graphs that produce correlated
particles from different color flux tubes extended in rapidity
with transverse separations less than the color-correlation
length, $1/Q_{s}$, where $Q_{s}$ is the saturation
scale~\cite{Dusling:2012iga,Dumitru:2010iy}. This picture results
in back-to-back particle correlations (i.e., $\Delta \phi \approx
0, \pi$) that extend long-range in pseudorapidity. With
$Q_{s} \approx 1\textnormal{--}2$~GeV, the transverse length scale
is 0.1--0.2~fm. Thus, the correlation should exist only among a
subset of the particles, and the correlation should be
predominantly back to back, resulting in a significant $v_{2}$ but
no significant $v_{3}$ or higher moments. The measurement of multiparticle correlations and
higher-order anisotropy coefficients in small
systems at RHIC and the LHC present a challenge for
these pictures. Recent research including additional diagrams indicate
that these features may be recovered at a qualitative level (see
Reference~\cite{Schenke:2017bog} for a summary).

A key test of the momentum domain pictures comes from the geometry
tests with \pau, \dau, and \heau collisions at RHIC. The momentum correlations
originate in a local domain size of order 0.2~fm, which is quite
small compared with the deuteron 
root-mean-square
%\hl{RMS}
%%% Comment
%\textbf{\textcolor{red}{[**AU: Does this stand for ``root-mean-square''?**]}}
%\textbf{\textcolor{blue}{[Yes; thank you, text changed to clarify.]}}
%%% Comment
diameter of 4.2~fm. Thus, a
natural prediction of these locally generated correlations is that
the signal should be smaller in \dau collisions than in
\pau\ collisions. In the \dau case, each domain is contained in only one local
hot spot originating from either the proton or neutron from
the deuteron, so the final correlation is diluted by the
particles emitted from the other uncorrelated hot spot. In
contrast, in the hydrodynamic picture the two hot spots evolve
and merge thus generating a larger $v_{2}$ in the \dau case. To date, no
successful explanation of this detailed geometry dependence from momentum domain calculations exists.

There are other momentum space explanations invoking color
reconnection~\cite{Ortiz:2013yxa}, radiating
antennas~\cite{Gyulassy:2014cfa}, and target field
anisotropy~\cite{Kovner:2010xk}. Explanations
invoking collectivity from interference~\cite{Blok:2017pui} and
color dipole orientation bias~\cite{Iancu:2017fzn} have recently been
put forward. In most of these pictures, the relation between small
systems at RHIC and the LHC is ignored; the elliptic, triangular, and quadrangular flow components have no natural connection (in
contrast to the case with initial geometry coupled with
hydrodynamics); and the relation of \pp to \pa to \nucnuc is ad
hoc or nonexistent. Interestingly there has been a recent
attempt to gauge the combined influence of initial-state
momentum-domain correlations and final-state
scattering~\cite{Greif:2017bnr} modeled via \bamps. 
The only way to advance these alternatives is to perform
%\hl{To advance these alternatives there is no replacement for} 
%%% Comment
%\textbf{\textcolor{red}{[**AU: OK\ to reword as ``The only way to advance these alternatives is to use...''?**]}}
%\textbf{\textcolor{blue}{[Thank you, we have modified the text and removed the highlighting.]}}
%%% Comment
comprehensive
calculations across energies, geometries, and observables.

\section{HYDRODYNAMIC DISCUSSION AND IMPLICATIONS}\label{HydroDiscuss}

The modern view of hydrodynamics is as an effective
theory that describes long-wavelength excitations of a system
after the microscopic degrees of freedom are integrated out. The
conserved charges of the theory in the simplest cases are simply
the system's four-momentum components, and the equations of motion
are the conservation equations $\partial_\mu T^{\mu\nu}=0$, where
$T^{\mu\nu}$ is the stress--energy tensor. The fields can be taken
as the fluid's four-velocity $u^\mu$ and the energy density
$\epsilon(T)$ (or alternatively the temperature $T$)\footnote{For
simplicity, we assume there are no other conserved charges with
associated fields.}. An equation of state specifying the pressure
$p = p(\epsilon)$ as a function of energy density suffices to
close this simple example. However, even for an ideal fluid
described by $T^{\mu\nu}_\mathrm{ideal} = (\epsilon+p)u^\mu u^\nu + p
g^{\mu\nu}$ the equations of motion are clearly nonlinear due to
both the form of $T^{\mu\nu}$ and the constraint $u_\mu u^\mu =
-1$ (here we take $c=1$ and use the so-called mostly plus metric
convention standardly used in the relativistic hydrodynamic
community).

Non-ideal behavior is typically separated from the ideal fluid
contribution to the stress--energy tensor:

\begin{equation}
T^{\mu\nu} = T^{\mu\nu}_\mathrm{ideal} + \pi^{\mu\nu} \quad .
\end{equation}
Working to first order in a derivative expansion, in the local
rest frame of the fluid defined by $u^\mu_\mathrm{LFR}=(1,0,0,0)$, the
spatial components of $\pi^{\mu\nu}$ are parameterized as 

\begin{equation}
\pi^{ij} = -\eta\ \Big(\frac{\partial u^i}{\partial
x^j}+\frac{\partial u^j}{\partial x^i}-\frac{2}{3}
\delta^{ij}\partial_k u^k\Big)-\zeta\ \partial_\mu u^\mu \quad ,
\end{equation}
where $\eta$ and $\zeta$ are the shear and bulk viscosities,
respectively. These expressions, while perfectly consistent with
the definitions of viscosity for nonrelativistic systems, lead to
acausal behavior in the relativistic equation of motion.
Mathematically, this is because the kernel $\sim
\exp[-x^2/(4\frac{\eta}{\epsilon+p}t)]$ is characteristic of
parabolic diffusion equation; physically the superluminal
behavior is encoded in the assumption that the system can react
instantaneously to a shear stress.

M\"{u}ller \cite{Muller:1967zza}, then later
Israel~\cite{Israel:1976tn} and
Israel-Stewart~\cite{Israel:1979wp}, developed 
a theory at second-order
%\hl{a theory second-order} 
%%% Comment
%\textbf{\textcolor{red}{[**AU: OK\ to reword as ``a theory at second order''?**]}}
%\textbf{\textcolor{blue}{[Thank you, we have modified the text and removed the highlighting.]}}
%%% Comment
in the gradient expansion that (at the linear level)
preserved causality through the introduction of a relaxation time
$\tau_\Pi$ for the nonequilibrium terms in the stress--energy
tensor. This parameter may be viewed as a regulator for the
effective theory~\cite{Spalinski:2016fnj} parameterizing the
non-hydrodynamic (damped) modes necessary to ensure causality. As
such, $\tau_\Pi$ is not an unbounded free parameter, as it must
satisfy $\tau_\Pi > \eta/(\epsilon+p)$ to ensure that linear
perturbations in the sound channel do not exceed the speed of
light~\cite{Baier:2007ix}. For a given system, it is necessary to
determine whether the hydrodynamic modes dominate its description
or whether there is a crucial dependence on the value of
$\tau_\Pi$ indicating that the so-called non-hydrodynamic modes,
namely the underlying physics of the regulator, are being studied
\cite{Denicol:2011fa,Denicol:2012cn}.

The necessity of the second-order term introduced by
M\"{u}ller-Israel-Stewart
%\hl{M\"{u}ller-Israel-Stewart}
%%% Comment
%\textbf{\textcolor{red}{[**AU: OK\ to change to ``M\"uller and Israel-Stewart'' (assuming Israel and Israel-Stewart are the same person)?**]}}
%\textbf{\textcolor{blue}{[Israel and Stewart are two separate authors. We have removed highlight and left text as is.]}}
%%% Comment
(MIS), together with the desire to apply
hydrodynamics in small hadronic systems, requires understanding
the order-by-order investigation of terms in the gradient
expansion. The relevant expansion parameter in weakly coupled
systems that admit a quasiparticle description is the Knudsen
number\footnote{Again, this is a simplified description; a more
general approach rooted in kinetic theory\cite{Denicol:2012cn}
allows for expansion in both $\mathrm{K}_N$ and the inverse Reynolds number
Re$^{-1} \sim |\pi^{\mu\nu}|/p $, contravening Landau's
expectation that these are essentially the same in relativistic
systems where hydrodynamics is applicable.} $\mathrm{K}_N \equiv \mfp/R$,
as noted in Landau's arguments in Section~\ref{Sec:History}.

Simple estimates of parton mean free paths, under the assumption
of the high parton density expected for a fully developed QGP,
provided only very modest support for the validity of a
hydrodynamical description in small hadronic systems:

\begin{equation}
\mfp \sim (2 \ \mathrm{fm} ) \left( \frac{T_0}{T} \right)^3 \left(
\frac{\sigma_1}{\sigma} \right), \label{Eq:mfp}
\end{equation}
where $T$ is the temperature of the plasma in MeV, $T_0 = 200$~MeV
(introduced to provide a scale; this is not the transition
temperature), $\sigma$ is the parton--parton cross section, and
$\sigma_1 = 1$~mb. Estimates such as Equation~\ref{Eq:mfp} have
been used to argue that for collisions of large nuclei with radii
$R \sim 6\mathrm{-}7$~fm, parton--parton cross sections no larger
than a few millibarns suffice to provide mean free paths
significantly smaller than the system size, ensuring $\mathrm{K}_N \leq
0.1$, down to temperatures of order 200~MeV or lower.

At the same time, it is clear that even for large nuclei, the
separation of scales between $\mfp$ and $R$ is at best an
order of magnitude. This observation leads directly to a fundamental question: What
is the smallest drop of liquid QGP? Two arguments suggest that a
size as small as a femtometer might be plausible. First, the
successes of hydrodynamics in describing the higher moments of the
flow harmonics suggested, circa 2010, that hydrodynamics was capable
of describing features in the data with sizes $\sim R /n$, where
$n$ is the order of the flow harmonic. Second, the small value of
$\eta/s$ inferred from the data argued that the QGP must be
strongly coupled, suggesting that the ballistic transport
assumptions used in the above expression for the mean free paths
is an overestimate. Nonetheless the observation of flow-like
features in $p+A$ and $p+p$ collisions was a surprising development
to most researchers.

Although the MIS theory was essential in
establishing the possibility of relativistic causal theory of
viscous hydrodynamics, it implemented only the minimal
second-order term needed to eliminate superluminal behavior. The
successes of the Little Bang model described in
Section~\ref{Sec:SMofHIC}, and the attendant interest in reliably
quantifying the key parameter ${\eta}/{s}$, motivated efforts
to systematically investigate all allowed second-order terms.

In two remarkable papers
submitted (independently) to the arXiv on the same day,
Baier et al. \cite{Baier:2007ix} and
Bhattacharyya et al. \cite{Bhattacharyya:2008jc}
used the gauge/gravity duality~\cite{Maldacena:1997re,Horowitz:2006ct} 
to not only investigate all five
allowed second-order terms in a conformal theory of relativistic
hydrodynamics but also calculate the magnitude of the
associated transport coefficients for a strongly coupled system
with the minimal value of ${\eta}/{s}$, in particular finding
\hbox{$\tau_\Pi = ({2-\ln 2})/{2\pi T} \approx {1.31}/{2\pi
T} \approx {0.21}/{T} $.} Romatschke~\cite{Romatschke:2009kr} extended this research to the
case of non-conformal hydrodynamics, in
which case there are 15 second-order terms, each with an
associated transport coefficient. Further efforts led to gradient
expansions to all orders in linearized theory~\cite{Bu:2014sia},
and only third-order in full theory~\cite{Grozdanov:2015kqa}.

These developments led to a greatly increased understanding of
relativistic hydrodynamics with important consequences for
evaluating its applicability in nuclear collisions in general, and
small systems in particular, in which gradients are large. There
is now a vast literature on the topic, drawing insights from
kinetic theory, linear response theory and the gauge/gravity
duality; for thorough and masterful reviews we refer the reader to
References~\citen{Florkowski:2017olj} and \citen{Romatschke:2017ejr}. Here we
summarize the important conclusions from those efforts:

\begin{enumerate}
\item The success of viscous relativistic hydrodynamics in describing the
bulk features, in particular the $v_n$s, does not necessarily
imply that the matter is near thermal equilibrium during its
hydrodynamic evolution. Rather, it is likely that high-energy
nuclear collisions remain out of equilibrium up to
hadronization~\cite{Romatschke:2016hle}. Obviously, by definition
hydrodynamics must be capable of addressing arbitrarily small
perturbations about local equilibrium, but recent research has shown
that this is also true in the case of momentum anisotropies of
order one.

\item The key condition for the applicability of hydrodynamics in small
systems is the dominance of hydrodynamics modes over
non-hydrodynamic modes. This appears to be a tautological
statement, but it can be put on a firm basis. Hydrodynamic modes
have dispersion relations satisfying $\lim_{|\mathbf{k}|\to 0}
\omega(\mathbf{k}) = 0$ consistent with the existence of conserved
charges central to the defining equations. Conversely,
non-hydrodynamic modes are those with finite imaginary values of
$\omega(\mathbf{k})$ as $\mathbf{k}$ goes to zero, indicative of
transient behavior not captured in the hydrodynamic gradient
expansion. Closely related to this observation is the divergence
of that gradient expansion \cite{Heller:2013fn,Buchel:2016cbj},
reflecting its inability to capture the non-hydrodynamic modes.
The non-hydrodynamic modes are an essential part of the early-time
dynamics necessary to insure consistency and/or causality, but
their late-time contributions to the dynamics should be small for
hydrodynamics to apply.

\item All studies to date~\cite{Florkowski:2017olj} indicate that the
first two to three orders of the gradient expansion provide a very
accurate description of a universal hydrodynamic attractor
behavior \cite{Heller:2015dha} for ${w \equiv \tau T_\mathrm{eff}(\tau) >
\sim 0.7}$~\cite{Heller:2011ju}, where $\tau$ is the time from the
initial collision and $T_\mathrm{eff}(\tau)$ is the {effective}
temperature at that time as determined from the energy or entropy
density. This is true even for systems that are grossly out of
equilibrium, that is, have momentum anisotropies of order one.

\item These considerations have led to the concept of a
{hydrodynamization} time,\footnote{In an interesting example
of confluence, this awkward but accurate construct first appeared
in an August 2013 paper on astrophysical plasmas
\cite{baranov2013effect} and in the heavy ion context in October of the same year \cite{Casalderrey-Solana:2013aba} and again
three weeks later \cite{Chesler:2013urd}.} in analogy to the
(not necessarily relevant) thermalization time, and defined as the
time when the hydrodynamic modes dominate the system's behavior.
All indications are that this time is of order $\tau_\mathrm{hydro} \sim
(0.5\mathrm{-}1.0)[{1}/({T_\mathrm{eff}(\tau)]}$ Note that at this time
first-order corrections to ideal hydrodynamics can still be large,
but the subsequent evolution is well described by viscous
relativistic hydrodynamics.

\item The requirement that the hydrodynamic modes dominate non-hydrodynamic
effects resulting from the second-order transport coefficient
$\tau_\Pi$ can be used to obtain a criterion on the smallest
system expected to exhibit hydrodynamic behavior. Three
semi-independent lines of
reasoning~\cite{Habich:2015rtj,Spalinski:2016fnj,vanderSchee:2017hsu}
led to the surprising conclusion
%%% Comment
%\textbf{\textcolor{red}{[**AU: Word missing here?**]}}
%\textbf{\textcolor{blue}{[Yes, thank you, we have modified the text.]}}
%%% Comment
that charged-particle rapidity densities
satisfying ${\mathrm{d}N_\mathrm{ch}}/{\mathrm{d}y} > 2$--$4$ suffice for a
valid description of system evolution using viscous relativistic
hydrodynamics.
\end{enumerate}

In summary, ample theoretical arguments developed over
the past decade suggest that viscous relativistic hydrodynamics
can be applied to describe particle production and flow in $p+p$ and
$p+A$ collisions at high energies. There is strong internal
consistency in this reasoning. The arguments (for the large part)
rely on strong coupling, which in turn implies a small value of
${\eta}/{s}$, which when used in hydrodynamics modeling
results in good agreement with the data. Similarly, the criterion
that $\mfp \sim {1}/{T}$ for minimal viscosity
systems~\cite{Danielewicz:1984ww} is echoed in the observation
\cite{Shuryak:2013ke,Basar:2013hea} that under these conditions a
plausible bound on the minimum system size $R$ for a hydrodynamic
description is $R \sim {1}/{T_\mathrm{eff}}$, which is supported by
numerical studies in a dual gravity
system \cite{Chesler:2015bba,Chesler:2016ceu}. Further support is
provided by the preservation of structure in the final state in
$A+A$ collisions up to at least $v_5$, which in effect are feature
sizes of order one-fifth those of the nuclear size.

\section{SUMMARY}
\begin{summary}[SUMMARY POINTS]
\begin{enumerate}
\item Small collision systems have
proved to be the perfect laboratory for studying the perfect-fluid behavior of quark-gluon plasma. 
As of early 2018,
%\hl{this writing}
%%% Comment
%\textbf{\textcolor{red}{[**AU: OK to replace with ``early 2018''?**]}}
%\textbf{\textcolor{blue}{[Yes, we have modified the text.]}}
%%% Comment
the field of relativistic heavy ion
physics is in the midst of a revolution in our understanding of
the conditions necessary for nuclear matter to behave as a
near-perfect fluid with bulk dynamics described by viscous
relativistic hydrodynamics. 
\item The revolution has been driven by the experimental 
observation of flow-like features in the collisions of small hadronic systems.
The theoretical insights are drawn from a broad range of studies ranging from 
relativistic kinetic theory to black-hole quasinormal modes in the context of the gauge/gravity duality. 
\item These studies have demonstrated that the hydrodynamization time, rather than the 
thermalization time, is the key parameter controlling the applicability of hydrodynamics to describe the evolution of systems, and that this criterion is valid even for systems far removed from thermal equilibrium. 
\item The insights derived from this ongoing work have greatly extended the regimes in which 
we can apply properly-formulated relativistic viscous hydrodynamics,
with implications for many-body strongly coupled systems in other fields of physics.
\end{enumerate}
\end{summary}
\newpage
\begin{issues}[OPEN QUESTIONS]
\begin{enumerate}
\item Are there alternatives to hydrodynamic modeling capable of simultaneously reproducing the experimental data from the geometry engineering of the initial state?
%for three nuclear systems? 
\item How can we understand the success of parton transport models that seemingly violate quantum mechanical limits yet reproduce flow-like features in small systems?
\item Are there experimental observables sensitive to the non-hydro modes? Can they be used to determine the associated relaxation parameters?
\item What is the smallest drop of QGP describable by relativistic viscous hydrodynamics?
\end{enumerate}
\end{issues}

% perfect fluid pervasive ubiquitous QCD
% small systems perfect laboratory for perfect fluid 

\section*{DISCLOSURE STATEMENT}

The authors are not aware of any affiliations, memberships,
funding, or financial holdings that might be perceived as
affecting the objectivity of this review.

\section*{ACKNOWLEDGMENTS}

We are pleased to acknowledge very useful comments from Jorge
Noronha, Jaki~Noronha-Hostler, Constantin Loizides, Krishna
Rajagopal, Paul~Romatschke and Bjoern Schenke. We also acknowledge work on the
graphics by Javier Orjuela-Koop. Some of the definitions that appear in the
margins have been taken from the heavy ion overview article \cite{Busza:2018rrf}
that appears in this volume. JLN and WAZ gratefully acknowledge funding
from the Division of Nuclear Physics of the US Department of
Energy under grants DE-FG02-00ER41152 and DE-FG02-86ER40281,
respectively.

%%% WAZ: Below three lines should be commented out before submission,
%%^ and contents of ultimate .bbl file swapped in below.
% \bibliographystyle{ar-style5}       %%% for ARNPS
\bibliographystyle{atlasnote}         %%% for arXiv
\bibliography{NagleZajcARNPS}
\end{document}